\documentclass{article}

\usepackage{arxiv}

\usepackage[utf8]{inputenc} % allow utf-8 input
\usepackage[T1]{fontenc}    % use 8-bit T1 fonts
\usepackage{hyperref}       % hyperlinks
\usepackage{url}            % simple URL typesetting
\usepackage{booktabs}       % professional-quality tables
\usepackage{amsfonts}       % blackboard math symbols
\usepackage{nicefrac}       % compact symbols for 1/2, etc.
\usepackage{microtype}      % microtypography
\usepackage{lipsum}		% Can be removed after putting your text content
\usepackage{graphicx}
\usepackage{natbib}
\usepackage{doi}
\usepackage{amsmath}

\usepackage{booktabs}%

\usepackage[section]{algorithm}
\usepackage{algpseudocode}

\DeclareMathOperator{\tr}{tr}
\DeclareMathOperator{\diag}{diag}
\newcommand{\vect}[1]{\boldsymbol{\mathbf{#1}}}
\newcommand{\dtheta}[1]{\frac{\partial}{\partial\theta_{#1}}}
\newcommand{\tran}{\mathrm{T}}
\newcommand{\Exp}{\mathbb{E}}
\newcommand{\Var}{\mathbb{V}}

\title{Parallel Selected Inversion for Space-Time~Gaussian~Markov Random Fields}

\date{September 12, 2023}	% Here you can change the date presented in the paper title
\author{ \href{https://orcid.org/0000-0001-7541-3537}{\includegraphics[scale=0.06]{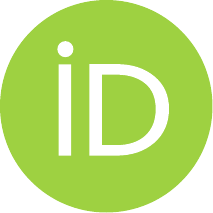}\hspace{1mm}Abylay Zhumekenov}\\
	\texttt{abylay.zhumekenov@kaust.edu.sa} \\
	\And
        \href{https://orcid.org/0000-0002-7063-2615}{\includegraphics[scale=0.06]{orcid.pdf}\hspace{1mm}Elias Krainski}\\
	\texttt{elias.krainski@kaust.edu.sa} \\
        \And
	\href{https://orcid.org/0000-0002-0222-1881}{\includegraphics[scale=0.06]{orcid.pdf}\hspace{1mm}H\r{a}vard Rue}\\
	\texttt{haavard.rue@kaust.edu.sa} \\	
	\and
	CEMSE Division\\
	King Abdullah University of Science and Technology\\
	Thuwal, 23955-6900, Saudi Arabia\\
}

\hypersetup{
pdftitle={Parallel Selected Inversion for Space-Time Gaussian Markov Random Fields},
pdfsubject={stat.CO},
pdfauthor={Abylay Zhumekenov},
pdfkeywords={latent Gaussian models, spatio-temporal, selected inverse, domain decomposition, distributed},
}

\begin{document}
\maketitle

\begingroup
\renewcommand\thefootnote{}\footnotetext{Published in \emph{Statistics and Computing} (2025). DOI: 10.1007/s11222-025-10747-y}
\addtocounter{footnote}{-1}
\endgroup

\begin{abstract}
Performing Bayesian inference on large spatio-temporal models requires extracting inverse elements of large sparse precision matrices for marginal variances, as well as estimating model hyperparameters. Although direct matrix factorizations can be used for the inversion, such methods fail to scale well for distributed problems when run on large computing clusters. On the contrary, Krylov subspace methods for the selected inversion have been gaining traction. We propose a parallel hybrid approach based on domain decomposition, which extends the Rao-Blackwellized Monte Carlo estimator for distributed precision matrices. Our approach exploits the strength of Krylov subspace methods as global solvers and efficiency of direct factorizations as base case solvers to compute the marginal variances and the derivatives required for hyperparameter estimation using a divide-and-conquer strategy. By introducing subdomain overlaps, one can achieve greater accuracy at an increased computational effort with little to no additional communication. We demonstrate the speed improvements and efficient hyperparameter inference on both simulated models and a massive US daily temperature data.
\end{abstract}

% keywords can be removed

% sections
\section{Introduction}\label{sec:1}

A need for an efficient parallel computation of selected elements of an inverse of a large sparse precision matrix often arises, when dealing with high-dimensional Bayesian spatio-temporal models obtained from stochastic partial differential equations (SPDE) \citep{lindgren2011explicit, lindgren2022spde}. The SPDE-derived models are extremely useful for many applications, ranging from disease mapping to climate analysis. Coupled with the integrated nested Laplace approximation (INLA) methodology for latent Gaussian models (LGM) \citep{rue2009approximate, rue2017bayesian}, the SPDE approach delivers the needed speed for large space-time applications. However, the growing amount of data as well as demand for high resolution modeling introduce additional computational challenges.

In particular, one of the major obstacles to a fast Bayesian inference on large LGMs is scalability of selectively inverting the posterior precision $\vect{Q}$ of the latent field. The selected inverse is needed for obtaining marginal variances $\diag(\vect{Q}^{-1})$ of the posterior latent field and for computing derivatives of the log-determinant with respect to certain hyperparameters via $\dtheta{j}\log|\vect{Q}|=\tr(\vect{Q}^{-1}\dtheta{j}\vect{Q})$. Taking into account that the derivative evaluation is a crucial component of any gradient-based learning method, it is essential that the chosen inversion method is fast and scalable. While being sparse (see Section \ref{sec:2}), the precision matrix could be prohibitively large for direct matrix factorizations. In this paper, we address this problem by taking a distributed hybrid approach for selected inversion.

Most conventional approaches for Bayesian inference, including INLA, rely heavily on direct matrix factorizations such as Cholesky decomposition $\vect{Q}=\vect{LL}^\tran$ for their computations. For historical reasons, these methods were initially implemented for shared memory devices, which are constrained by memory and number of cores. There has been some recent work to parallelize INLA to run on several cluster nodes to achieve greater scalability \citep{fattah2022approximate, gaedke2023parallelized, gaedke2024integrated}. However, the implementations are limited to parallel function evaluations of the posterior, where each node stores the full precision matrix with its Cholesky factor and parallelizes matrix operations using OpenMP. The proposed methods cannot tackle larger problems, where even storing and accessing the matrix becomes challenging, let alone obtaining results in reasonable time.

The issue is handled by distributing the matrix across multiple shared memory devices or cluster nodes. Although parallel direct techniques for the inversion do exist, the scalability of such methods leaves a gap for improvement. For example, a great boost in speed is achieved for reordered 3D problems by using Takahashi equations for the selected inversion \citep{erisman1975computing}, instead of naively solving for the entire inverse. An additional speed can be gained by parallelizing the numerical factorization step using directed acyclic graphs (DAG) for task-scheduling \citep{liu1986computational}. Nevertheless, these tools are merely optimizations to an inherently serial algorithm. What prevents Cholesky to scale on large clusters is high inter-node latency and low computation-to-communication ratio of the factorization \cite{li2003superlu_dist}.

On the other hand, Krylov methods for the selected inversion have been gaining more attention recently. Krylov subspace methods is a large class of projection methods for sparse systems, which have low memory requirements, do not perform heavy matrix operations and compute only sparse matrix-vector products, which are very easy to implement and evaluate in parallel \citep{saad2003iterative}. This allows Krylov solvers to scale extremely well in a distributed setting, and they have been extensively used for solving large problems in PDE and lately in statistics. They have also been used for large scale sampling from GMRF \citep{schneider2003krylov, simpson2008fast, papandreou2010gaussian, parker2012sampling, simpson2013scalable, chow2014preconditioned}, and more importantly, for computing certain inverse elements of large sparse matrices \citep{hutchinson1989stochastic, tang2012probing, siden2018efficient}. 

However, most of the papers do not go beyond shared memory implementation or do not consider the case, when it is prohibitive to store the precision matrix and its Cholesky factor on a single machine. In this paper, we will present an extension to the Rao-Blackwellized Monte Carlo algorithm from \citet{siden2018efficient} for large distributed matrices for estimating hyperparameters of diffusion based spatio-temporal models \citep{lindgren2020diffusion}. Our method can be naturally derived using a divide-and-conquer paradigm, which connects it to many existing domain decomposition techniques. The proposed solution is fast, communication friendly, highly parallel and most importantly, distributed.

The rest of the paper is organized as follows. In Section \ref{sec:2}, we will give a general background for existing inversion methods and Bayesian inference on non-separable spatio-temporal models. In Section \ref{sec:3}, we demonstrate how Krylov methods can be used to compute the posterior parameters of a latent model. In particular, we will introduce a recursive algorithm for approximating the posterior marginal variance, and reformulate it as a parallel algorithm for overlapping domains. A simulation study and an application to a real world example will be given in Section \ref{sec:4}.

\section{Background}\label{sec:2}

\subsection{Stochastic partial differential equations approach}

For the past decade, the SPDE approach has been widely used for modeling of spatio-temporal Gaussian processes \citep{lindgren2011explicit}\citep{lindgren2022spde}. The approach  presents itself as a fast and scalable alternative to covariance matrix based methods and relies on a Gaussian Markov random field (GMRF) approximation to a continuously defined random process. The main computational advantage comes from the Markov property, which results in a sparse precision structure, allowing one to employ more efficient algorithms. 

The spatial Whittle-Mat\'ern fields serve as a starting point for constructing spatio-temporal models and many other extension of the SPDE approach. The spatial field is defined as a stationary solution $u(\vect{s})$ to the equation
\begin{equation}\label{eq:spde:1}
    \tau(\kappa^2-\Delta)^{\alpha/2}u(\vect{s}) = \mathcal{W}(\vect{s}),
\end{equation}
where $\tau$ is a precision parameter, $\kappa$ is inversely related to the range, and $\alpha$ is a positive integer related to the smoothness of the field.
It has been shown in \cite{whittle1954stationary} and \cite{whittle1963stochastic} that the solution $u(\vect{s})$ has the Mat\'ern covariance function. Generally, the equation could be solved using a weak formulation and a finite element discretization of the spatial domain. The approximate solution is expressed as a linear combination $\Tilde{u}(\vect{s})=\sum u_i\psi_i(\vect{s})$ of basis functions $\psi_i$, which are usually chosen to be piecewise linear and have a compact support. For $\alpha=2$, one would solve for the solution $\vect{u}$ the following linear system
\begin{equation}\label{eq:spde:2}
    \tau(\kappa^2\vect{C}+\vect{G})\vect{u} = \vect{z}.
\end{equation}
Here $\vect{C}$ and $\vect{G}$ are sparse matrices better known as mass and stiffness matrices, respectively. Now we are interested in the precision operator of the discretized solution, let us denote it as $\vect{Q}_{s}$. The calculations would yield $\vect{Q}_{s} = (\kappa^2\vect{C}+\vect{G})\vect{C}^{-1}(\kappa^2\vect{C}+\vect{G})$
for $\alpha=2$ and
\begin{equation}\label{eq:spde:4}
    \vect{Q}_{s} = \tau^2\vect{C}^{1/2}\left( \kappa^2\vect{I}+\vect{C}^{-1/2}\vect{GC}^{-1/2} \right)^\alpha\vect{C}^{1/2}.
\end{equation}
for the general case. It must be noted that the precision matrix is not sparse, unless the mass matrix $\vect{C}$ is made diagonal. This can be done using a procedure, called ``mass lumping'', which is common in FEM applications \cite{chen1985lumped}. Then the solution vector $\vect{u}$ has a distribution $\mathcal{N}(\vect{0},\vect{Q}_{s}^{-1})$, where the precision matrix is sparse, and hence, $\vect{u}$ is a GMRF.

\subsection{Non-separable space-time models}

From this point, the precision matrix in Equation \eqref{eq:spde:4} can be used as a building block for separable or non-separable spatio-temporal extensions. The separable case is discussed in \citet{lindgren2015bayesian}. A common way to construct a spatio-temporal precision matrix for a separable model is through a Kronecker product $\vect{Q}_{u}=\vect{Q}_{t}\otimes\vect{Q}_{s}$ between a temporal precision $\vect{Q}_{t}$ and a spatial $\vect{Q}_{s}$ derived above. However, separability is a strong assumptions for most practical applications. \citet{lindgren2020diffusion} discusses several attempts for non-separable extensions and introduces a new class of diffusion-based non-separable space-time processes.

The diffusion models can be defined by considering the spatial model first:
\begin{equation}\label{eq:spde:5}
    \gamma_e\mathcal{L}_s^{\alpha_e/2}v(\vect{s}) = \mathcal{W}(\vect{s}),
\end{equation}
where $\mathcal{L}_s=\gamma_s^2-\Delta$. The authors then construct a spatio-temporal process using the SPDE
\begin{equation}\label{eq:spde:6}
    \left( -\gamma_t^2\frac{\partial^2}{\partial t^2}+\mathcal{L}_s^{\alpha_s} \right)^{\alpha_t/2} u(\vect{s},t) = d\mathcal{E}(\vect{s},t),
\end{equation}
where $(\vect{s},t)\in\mathcal{D}\times\mathbb{R}$, and $d\mathcal{E}(\vect{s},t)$ is defined as a Gaussian noise, which is white in time, but is correlated in space according to \eqref{eq:spde:5}. The parameters $\gamma_t,\gamma_s,\gamma_e$ are scale parameters, while $\alpha_t,\alpha_s,\alpha_e$ together determine the spatial and temporal smoothness, as well as the separability of the space-time model. The simplest non-separable case is when $(\alpha_t,\alpha_s,\alpha_e) = (1,2,1)$ and $d=2$, referred to as ``critical diffusion''. The table with different parameter values and models can be found in \citep{lindgren2020diffusion}. These values are usually fixed to positive integers for the SPDE approach, this ensures the desired Markov property of the weights $\vect{u}$. Otherwise, rational approximations could be used for fractional values of $\alpha$ \citep{bolin2020rational}.

The corresponding finite element approximation to the solution of \eqref{eq:spde:6} is expressed through the general Kronecker product basis expansion
\begin{equation}\label{eq:spde:7}
    u(\vect{s},t) = \sum_{i=1}^{n_s}\sum_{j=1}^{n_t}u_{ij}\psi_i(\vect{s})\varphi_j(t),
\end{equation}
where $\psi_i$ and $\varphi_j$ are basis functions defined in space and time, respectively. Similar to the spatial case, piecewise linear basis functions with compact support produce a sparse system of linear equations. From there, the precision matrix of the flattened solution vector $\vect{u}=(u_{11},u_{21},\dots,u_{n_sn_t})$ can be computed as
\begin{equation}\label{eq:spde:8}
    \vect{Q}_{u} = \gamma_e^2 \sum_{k=0}^{2\alpha_t} \gamma_t^k\vect{J}({\alpha_t,k/2}) \otimes \vect{K}({\alpha_s(\alpha_t-k/2)+\alpha_e}),
\end{equation}
for some sparse symmetric $\vect{J}$ matrices coming from a temporal discretization and previously derived precision matrices $\vect{K}$ of a spatial process. The variance and the range of the field are controlled jointly by hyperparameters $\gamma_t,\gamma_s,\gamma_e$, which are not interpretable when considered separately. Authors propose a reparametrization to practical spatial range, practical temporal range and marginal variance, $r_s, r_t, \sigma^2$. A proper mapping between $\gamma_t,\gamma_s,\gamma_e$ and $r_s, r_t, \sigma^2$ for different domains and manifolds can also be found in the paper \cite{lindgren2020diffusion}.

The resulting sum of Kronecker products is sparse and inherits the block structure of corresponding temporal matrices. The in-block structure comes from the spatial mesh and can be considered arbitrary in general. While a single Kronecker product can be decomposed into a product of Cholesky factorizations, the fact that the precision matrix is expressed as a sum does not allow to reuse the factors. In general, a factorization of a sum of matrices cannot be obtained unless the matrix is assembled. On the other hand, individual factors can be used for sampling as in \citet{papandreou2010gaussian}. In our case, one would need to decompose each term as
\begin{equation}\label{eq:spde:9}
\begin{aligned}
    \vect{J}\otimes\vect{K} 
    & = (\vect{L}\vect{L}^\tran)\otimes(\vect{R}\vect{R}^\tran) \\
    & = (\vect{L}\otimes\vect{L})(\vect{R}\otimes\vect{R})^\tran,
\end{aligned}
\end{equation}
with $\vect{L}$, $\vect{R}$ being Cholesky factors of $\vect{J}$, $\vect{K}$ respectively. However, the $\vect{J}$ matrices might be positive semi-definite, meaning that there is no unique Cholesky factorization and the algorithm will break down. Even if one finds such a decomposition, the proposed solution must be as robust and efficient as Cholesky for arbitrary stiffness matrices.

\subsection{Latent Gaussian models}

The SPDE approach is a very useful tool for setting physics informed priors for latent Gaussian models (LGM).
LGM is a large class of three stage hierarchical Bayesian model that encompasses many important statistical models, including space-time diffusion models we discussed above. The three levels are observations $\vect{y}\in\mathbb{R}^{n_y}$, the latent Gaussian field $\vect{x}=(\vect{u},\vect{\beta})\in\mathbb{R}^{n}$ comprised of latent processes $\vect{u}\in\mathbb{R}^{n_u}$ and fixed effects $\vect{\beta}\in\mathbb{R}^{n_{\beta}}$ ($n=n_u+n_{\beta}$), and hyperparameters $\vect{\theta}\in\mathbb{R}^{n_{\theta}}$ which control the latent field and the likelihood. For the sake of simplicity and as a starting point, this paper considers the case when the likelihood is Gaussian and the hyperparameters are fixed. It is easy to generalize to unknown hyperparameters, which we will do at the end of Section \ref{sec:3}.
\begin{equation}\label{eq:lgm:1}
\begin{aligned}
    \vect{y}|\vect{x},\vect{\theta} &\sim \mathcal{N}(\vect{Ax},\vect{Q}_{y}^{-1})\\
    \vect{x}|\vect{\theta} &\sim \mathcal{N}(\vect{0},\vect{Q}_{x}^{-1})\\
    \vect{\theta} &= \vect{\theta}_0 \text{(fixed)}
\end{aligned}
\end{equation}
Here, the matrix $\vect{A}=[\vect{A}_{u}\;\vect{A}_{\beta}]\in\mathbb{R}^{n_y\times n}$ consists of a projection matrix $\vect{A}_{u}\in\mathbb{R}^{n_y\times n_u}$ for the FEM solution (see \cite{lindgren2011explicit}) and $\vect{A}_{\beta}\in\mathbb{R}^{n_y\times n_{\beta}}$ is a thin matrix of covariates. The matrix $\vect{Q}_{x}\in\mathbb{R}^{n\times n}$ is a block diagonal prior precision matrix of $\vect{x}$, with the prior precisions $\vect{Q}_{u}\in\mathbb{R}^{n_u\times n_u}$ and $\vect{Q}_{\beta}\in\mathbb{R}^{n_{\beta}\times n_{\beta}}$ of the spatio-temporal field and the fixed effects respectively, placed on the diagonal. $\vect{Q}_{y}\in\mathbb{R}^{n_y\times n_y}$ is the precision of the observations. In general, the latent field will include other random effects as well, but we will consider a simpler setting in this paper.

The objective is to obtain the posterior marginals for the unknown latent variables $\vect{x}$. In our case, the joint posterior density corresponds to the full conditional $\pi(\vect{x}|\vect{y})=\pi(\vect{x}|\vect{y},\vect{\theta})$ and can be obtained analytically:
\begin{equation}\label{eq:lgm:2}
    \vect{x}|\vect{y} \sim \mathcal{N}(\vect{\mu},\vect{Q}^{-1}),
\end{equation}
where $\vect{Q}=\vect{Q}_{x}+\vect{A}^\tran\vect{Q}_{y}\vect{A}$ is the posterior precision and $\vect{\mu}=\vect{Q}^{-1}\vect{A}^\tran\vect{Q}_{y}\vect{y}$ is the posterior mean. Then the marginal densities can be expressed as
\begin{equation}\label{eq:lgm:3}
    x_i|\vect{y} \sim \mathcal{N}(\mu_i,\sigma_i^2),
\end{equation}
where one needs to somehow compute $(\sigma_1^2,\dots,\sigma_{n}^2)=\diag(\vect{Q}^{-1})$. While the mean can be computed rather easily using iterative methods, extracting diagonal elements requires selectively inverting $\vect{Q}$. We note that the selected inverse can also be used for computing $\tr(\vect{Q}^{-1}\dtheta{}\vect{Q})$ when forming the gradient. In fact, one only needs to sum over an element-wise product 
\begin{equation}\label{eq:lgm:4}
    \tr\Big(\vect{Q}^{-1}\dtheta{}\vect{Q}\Big) = \sum_{ij}\Big(\vect{Q}^{-1}\Big)_{ij}\Big(\dtheta{}\vect{Q}\Big)_{ij}
\end{equation}
between selected inverse and the matrix derivative. We will now revisit existing methods for GRMFs in more detail. 

\subsection{Existing methods}

Most existing methods on large precision matrices rely on Krylov methods to estimate the selected entries of the inverse. Krylov subspace methods are iterative algorithms designed to efficiently solve large sparse linear systems $\vect{Q}\vect{x}=\vect{b}$. Given a matrix $\vect{Q}$ and vector $\vect{b}$, Krylov methods approximate the solution in a subspace $\mathcal{K}_m(\vect{Q},\vect{b}) = \text{span}\{\vect{b}, \vect{Q}\vect{b}, \dots, \vect{Q}^{m-1}\vect{b}\}$ of size $m$. These methods require only $\mathcal{O}(mn)$ storage and avoid the high computational cost of direct solvers. Their efficiency comes from simple vector operations and minimal communication, making them well-suited for large-scale problems. Additionally, their reliance on matrix-vector products makes them naturally parallelizable, further enhancing performance.  

Hutchinson's stochastic estimator \citep{hutchinson1989stochastic}, given in Equation \eqref{eq:intro:2}, is one of the oldest and well known methods for estimating the diagonal of the inverse. The estimator solves a sequence of random zero-centered right hand sides $\vect{z}^{(k)}$ using a Krylov method, and aggregates the result by embarassingly parallel element-wise vector operations. It is an unbiased estimator of the diagonal of the inverse and has a Monte Carlo like convergence properties, but is noisy for small sample sizes $n_k$. 
\begin{equation}\label{eq:intro:2}
\begin{aligned}
    \diag(\vect{Q}^{-1}) 
    &\approx \left[ \sum_{k=1}^{n_k}\vect{z}^{(k)}\odot\vect{Q}^{-1}\vect{z}^{(k)} \right] \\
    &\qquad\oslash \left[ \sum_{k=1}^{n_k}\vect{z}^{(k)}\odot\vect{z}^{(k)} \right],
\end{aligned}
\end{equation}

A probing method by \citet{tang2012probing} was derived as a deterministic version of the stochastic estimator for a sparsified inverse, see Equation \eqref{eq:intro:3}. Although the inverse is dense in general, it can be approximated by a sparse matrix with a non-zero pattern of $\vect{Q}^{n_q}$ for some power $n_q$. In this case, one could solve for several diagonal entries simultaneously using a single right hand side $\vect{z}^{(k)}$ consisting of ones at corresponding entries. The idea is that if the rows of $\vect{\Sigma}$ decay exponentially, we can find a set of variables, which are almost independent. The authors generate the vectors $\vect{z}^{(k)}$ using a greedy coloring technique. The number of vectors will typically depend on the order $n_q$ of the sparseness, which can be small for diagonally dominant matrices, but can grow large if it is not the case. For SPDE precision matrices, $n_q$ could be somehow associated with the correlation range, meaning that $n_q$ could be as big as $n$ in cases where the field is extremely correlated.
\begin{equation}\label{eq:intro:3}
    \diag(\vect{Q}^{-1}) \approx \diag(\vect{Q}^{-1}\vect{ZZ}^\tran) \oslash \diag(\vect{ZZ}^\tran).
\end{equation}

A different approach was taken by \citet{siden2018efficient} and their Rao-Blackwellized Monte Carlo (RBMC) estimator. In essence, authors improve the error of a sample variance estimator by conditioning on set of neighboring elements. The marginal variances can be computed by utilizing the law of total variance. The estimator is also unbiased and has similar convergence properties, but borrows information from the precision matrix itself. The basic version is given in Equation \eqref{eq:intro:4}, where $\vect{x}_{-i}^{(k)}$ denotes a sample of the latent field with its $i$-th element / row removed. 
\begin{equation}\label{eq:intro:4}
\begin{aligned}
    \Var(x_i) 
    & = \Exp[\Var(x_i|\vect{x}_{-i})] + \Var[\Exp(x_i|\vect{x}_{-i})] \\
    & = \Exp[q_{ii}^{-1}] + \Var[-q_{ii}^{-1}\vect{Q}_{i,-i}\vect{x}_{-i}] \\
    & \approx q_{ii}^{-1} + \frac{1}{n_k}\sum_{k=1}^{n_k}q_{ii}^{-1}\vect{Q}_{i,-i}\vect{x}_{-i}^{(k)},
\end{aligned}
\end{equation}
The authors also derive a blocked and interface versions of the algorithm. The main idea behind the interfaces is similar to creating a buffer region that absorbs effects from the conditioning set. At an increased computational cost, the estimators outperform Hutchinson's stochastic estimator in terms of accuracy.

One of the assumptions that makes the results in \citet{siden2018efficient} computationally viable is a simplified strategy for sampling from the GMRF. The precision matrix in the paper can be represented in the form $\vect{Q}=\vect{L}\vect{L}^\tran+\vect{R}\vect{R}^\tran$. Then, as suggested by \citet{papandreou2010gaussian}, one can easily sample from the posterior by solving $\vect{Q}\vect{u}=\vect{L}\vect{z}+\vect{R}\vect{w}$ using the conjugate gradient (or any Krylov) method, where $\vect{z}$ and $\vect{w}$ are standard Gaussian samples. However, this cannot always be done for matrices obtained from the FEM discretization of SPDE models. For instance, precision matrices for non-separable diffusion-based spatio-temporal models can be expressed as a sum of kronecker products \citep{lindgren2020diffusion}. However, some terms are positive semi-definite, and in such cases, the Cholesky algorithm breaks due to a zero pivot.

There are other papers that explore fast and efficient sampling of GMRFs using Krylov solvers \citep{schneider2003krylov, simpson2008fast, papandreou2010gaussian, parker2012sampling, simpson2013scalable, chow2014preconditioned}. Most of the methods rely on Lanczos tridiagonalization of the operator $\vect{Q}$ to build a low-rank approximation of the covariance matrix. However, the produced samples are of low quality and may not converge to the desired distribution. \citet{chow2014preconditioned} proposed a better method for obtaining high quality samples by approximating $\vect{Q}^{-1/2}\vect{z}$. Although each sample is obtained using a low rank Krylov approximation, the realized covariance converges to its full rank counterpart. 

In the next section, we will present our approach to solving the inversion problem with heavy emphasis on the scalability and parallel implementation. We will reformulate the results from \citet{siden2018efficient} as a domain decomposition method for distributed problems and use the preconditioned Krylov sampling from \citet{chow2014preconditioned} for parallel precision matrices.

\section{Main results}\label{sec:3}

\subsection{Computing the mean}

The posterior mean is a solution to a sparse linear system of equations
\begin{equation}\label{eq:mean:1}
    \vect{\mu} = \vect{Q}^{-1}(\vect{A}^\tran\vect{Q}_{y}\vect{y}).
\end{equation}
Preconditioned Krylov methods are an efficient way to solve such systems concurrently, but the efficiency will heavily depend on the spectrum of the operator $\vect{Q}$, its sparseness and load balancing, i.e. how the system was partitioned. Typically, large matrices are distributed according to row partitions, where each process owns only several rows of the global matrix. Then the matrix-vector products can be evaluated at each processor in parallel:
\begin{equation}\label{eq:mean:2}
    \vect{Q}\vect{v}^{(j)} =
    \begin{bmatrix}
        \vect{Q}_{\text{proc }1} \\
        \vdots \\
        \vect{Q}_{\text{proc }p}
    \end{bmatrix} \vect{v}^{(j)} =
    \begin{bmatrix}
        \vect{Q}_{\text{proc }1}\vect{v}^{(j)} \\
        \vdots \\
        \vect{Q}_{\text{proc }p}\vect{v}^{(j)}
    \end{bmatrix}.
\end{equation}

However, the block structure of the posterior precision matrix $\vect{Q}$ is not well suited for Krylov solvers. The structure is given below in Equation \eqref{eq:mean:3}. The prior $\vect{Q}_{x}$ is a block diagonal matrix consisting of a large sparse precision $\vect{Q}_{u}$ defined through SPDE, and a small diagonal $\vect{Q}_{\beta}$ corresponding to fixed effects. The matrix $\vect{Q}_{y}=\tau_y\vect{I}$ is a diagonal precision for observed data.
\begin{equation}\label{eq:mean:3}
\begin{aligned}
    \vect{Q}
    &= \vect{Q}_{x}+\vect{A}^\tran\vect{Q}_{y}\vect{A} \\
    &=
    \begin{bmatrix}
        \vect{Q}_{u} & \vect{0} \\
        \vect{0} & \vect{Q}_{\beta} \\
    \end{bmatrix} + 
    \begin{bmatrix}
        \vect{A}_{u}^\tran \\
        \vect{A}_{\beta}^\tran
    \end{bmatrix}
    \begin{bmatrix}
        \vect{Q}_{y}
    \end{bmatrix}
    \begin{bmatrix}
        \vect{A}_{u} & \vect{A}_{\beta}
    \end{bmatrix} \\
    &=
    \begin{bmatrix}
        \vect{Q}_{u} + \vect{A}_{u}^\tran\vect{Q}_{y}\vect{A}_{u} &
        \vect{A}_{u}^\tran\vect{Q}_{y}\vect{A}_{\beta} \\
        \vect{A}_{\beta}^\tran\vect{Q}_{y}\vect{A}_{u} &
        \vect{Q}_{\beta} + \vect{A}_{\beta}^\tran\vect{Q}_{y}\vect{A}_{\beta}
    \end{bmatrix} \\
    &=:
    \begin{bmatrix}
        \vect{Q}_{uu} & \vect{Q}_{u\beta} \\
        \vect{Q}_{\beta u} & \vect{Q}_{\beta\beta}   
    \end{bmatrix}
\end{aligned}
\end{equation}

The INLA implementation uses the matrix form above for its computations, where $\vect{A}_{\beta}$, and hence $\vect{Q}_{\beta u}$ are dense blocks \citep{rue2009approximate, rue2017bayesian, van2023new}. However, when assembling the matrix in parallel, it is important that matrix row partitions are equally sparse and balanced. The processor owning the $\vect{Q}_{\beta u}$ dense block will spent more time for performing more floating point operations when computing matrix-vector products $\vect{Q}\vect{v}^{(j)}$. This will result in overall algorithm slowdown, since Krylov solvers require synchronizations every or every few iterations.

Therefore, it is preferable to separate the SPDE part from the fixed effects. This can be done by conditioning on fixed effects $\vect{\beta}$, which gives the conditional precision operator $\vect{Q}_{u|\beta}=\vect{Q}_{uu}$. This lets us work with a sparser and more homogeneous object $\vect{Q}_{uu}$, for which the Krylov solver will be efficient. In addition, picking a preconditioner becomes easier, as we deal with the SPDE precision only, and the covariates have no effect on the spectrum anymore. The rest is handled by appropriate dense linear algebra routines. For the equation \eqref{eq:mean:1}, it means solving two systems
\begin{equation}\label{eq:mean:4}
    \begin{bmatrix}
        \vect{Q}_{uu} & \vect{Q}_{u\beta} \\
        \vect{Q}_{\beta u} & \vect{Q}_{\beta\beta}
    \end{bmatrix}
    \begin{bmatrix}
        \vect{\mu}_{u} \\
        \vect{\mu}_{\beta}
    \end{bmatrix} =
    \begin{bmatrix}
        \vect{A}_{u}^\tran\vect{Q}_{y}\vect{y} \\
        \vect{A}_{\beta}^\tran\vect{Q}_{y}\vect{y}
    \end{bmatrix}.
\end{equation}
By rearranging and substituting the terms, we get
\begin{equation}\label{eq:mean:5}
\begin{aligned}
    \vect{\mu}_{\beta} &= \vect{S}^{-1} \left((\vect{A}_{\beta}^\tran\vect{Q}_{y}\vect{y}) - \vect{Q}_{\beta u}\vect{Q}_{uu}^{-1}(\vect{A}_{u}^\tran\vect{Q}_{y}\vect{y})\right) \\
    \vect{\mu}_{u} &= \vect{Q}_{uu}^{-1} \left((\vect{A}_{u}^\tran\vect{Q}_{y}\vect{y}) - \vect{Q}_{u\beta}\vect{\mu}_{\beta}\right),
\end{aligned}
\end{equation}
where $\vect{S}=\vect{Q}_{\beta\beta}-\vect{Q}_{\beta u}\vect{Q}_{uu}^{-1}\vect{Q}_{u\beta}$ is the Schur complement of $\vect{Q}_{uu}$ in $\vect{Q}$, and also $\Var(\vect{\beta})=\vect{S}^{-1}$. We note that the posterior mean requires only $n_{\beta}+2$ solves with a sparse $\vect{Q}_{uu}$, one solve with a small dense $\vect{S}$, and a few additional yet cheap matrix-vector and matrix-matrix products.

\subsection{Computing the variance}

We can find the marginal variances by conditioning on fixed effects as
\begin{equation}\label{eq:var:1}
\begin{aligned}
    \diag(\Var(\vect{u})) 
    & = \diag(\vect{Q}_{uu}^{-1}) \\
    & + \diag(\vect{Q}_{uu}^{-1}\vect{Q}_{u\beta}\Var(\vect{\beta})\vect{Q}_{\beta u}\vect{Q}_{uu}^{-1}),
\end{aligned}
\end{equation}
where $\Var(\vect{\beta})=\vect{S}^{-1}$ can be computed from the Schur complement as mentioned earlier. We see that the only additional computation is $\diag(\vect{Q}_{uu}^{-1})$, as everything else is already computed. We will use Rao-Blackwellized Monte Carlo approach discussed in \citet{siden2018efficient} to extract the diagonal of the posterior covariance $\diag(\Var(\vect{u}))$ in parallel. The resulting algorithm can be derived both recursively and non-recursively. We will define an algorithm on $\vect{Q}$, but we can apply the algorithm directly on $\vect{Q}$ or on its submatrices, for example $\vect{Q}_{uu}$.

\subsubsection{Recursive RBMC}

Let us ease the notation and denote $\vect{x}:=\vect{x}|\vect{y}$, we will revert to the conditioning notation whenever it is necessary to be unambiguous. Then let $\vect{x}$ be a GMRF with respect to an undirected graph $G=(\Omega,E)$, with sets of vertices $\Omega=\{1,\dots,n\}$ and edges $E=\{(i,j)\in\Omega\times\Omega: x_i\perp x_j|x_{- ij}\}$, similar to the definition in \cite{rue2005gaussian}. A separating set $S$ is a set of vertices that disconnects the graph into two disjoint subgraphs with vertices $\Omega_1$ and $\Omega_2$. That is, there exists no $(i,j)\in \Omega_1\times\Omega_2$ such that $(i,j)\in E$. In probabilistic terms, $\vect{x}_{\Omega_1}^{}$ and $\vect{x}_{\Omega_2}^{}$ are conditionally independent, given $\vect{x}_{S}^{}$. 

Now, let $\vect{Q}_{\Omega_1\Omega_1}$ and $\vect{Q}_{\Omega_2\Omega_2}$ be principal submatrices obtained by selecting rows and columns indexed by $\Omega_1$ and $\Omega_2$, respectively. We form $\vect{Q}_{\Omega_1S}^{}$, $\vect{Q}_{\Omega_2S}^{}$ and $\vect{x}_{S}^{}$ in a similar fashion. Using the law of total variance, we obtain
\begin{equation}\label{eq:var:2}
\begin{aligned}
    \Var(\vect{x}_{\Omega_1}^{})
    &= \Exp(\Var(\vect{x}_{\Omega_1}^{}|\vect{x}_{S}^{})) + \Var(\Exp(\vect{x}_{\Omega_1}^{}|\vect{x}_{S}^{}))\\
    % &= \Exp(\vect{Q}_{\Omega_1\Omega_1}^{-1}) + \Var(-\vect{Q}_{\Omega_1\Omega_1}^{-1}\vect{Q}_{\Omega_1S}^{}\vect{u}_{S}^{})\\
    &= \vect{Q}_{\Omega_1\Omega_1}^{-1} + \Var(\vect{Q}_{\Omega_1\Omega_1}^{-1}\vect{Q}_{\Omega_1S}^{}\vect{x}_{S}^{})
\end{aligned}
\end{equation}
and
\begin{equation}\label{eq:var:3}
    \Var(\vect{x}_{\Omega_2}^{}) 
    = \vect{Q}_{\Omega_2\Omega_2}^{-1} + \Var(\vect{Q}_{\Omega_2\Omega_2}^{-1}\vect{Q}_{\Omega_2S}^{}\vect{x}_{S}^{})
\end{equation}
for the other half. We can see that the only common term in formulas \eqref{eq:var:2} and \eqref{eq:var:3} is $\vect{x}_{S}^{}$. That is, we can compute these two separately and in parallel, given samples of $\vect{x}_{S}^{}$ to evaluate the variance term.

\begin{figure}[ht]
    \centering
    \includegraphics[width=0.7\linewidth]{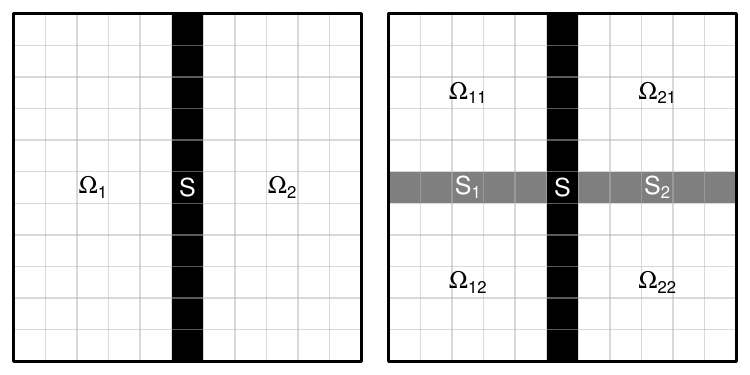}
    \caption{Two iterations of the recursive dissection algorithm}
    \label{fig:var:1}
\end{figure}

Moreover, computing the first term $\vect{Q}_{\Omega_1\Omega_1}^{-1}$ or $\vect{Q}_{\Omega_2\Omega_2}^{-1}$ is exactly the same problem as computing $\vect{Q}^{-1}$, but half the size. We could repeat the procedure for $\vect{x}_{\Omega_1}^{}|\vect{x}_{S}^{}$ to produce a partition $\Omega_{11},\Omega_{12},S_1$, subsets $\vect{x}_{\Omega_{11}}$, $\vect{x}_{\Omega_{12}}$ and a conditioning set $\vect{x}_{S\cup S_1}$. We could keep splitting until we reach a dimension small enough for direct methods. This allows a recursive specification of the inversion problem, a sketch of the algorithm is given in Algorihtm \ref{alg:var:1}, as well as an illustration of the divide-and-conquer strategy in Figure \ref{fig:var:1}.

\begin{algorithm}[ht]
\caption{Recursive RBMC}\label{alg:var:1}
\textbf{Input}: $\vect{Q}$, $\vect{x}_{S_0}$
\begin{algorithmic}
    \If{$\vect{Q}$ is small}
        \State return $\vect{Q}^{-1}$
    \EndIf
    \State $\vect{\Sigma} = \vect{0}$
    \State Partition the graph into $\Omega_1,\Omega_2,S$
    \State Set $S:=S\cup S_0$
    \State $\vect{\Sigma}_{\Omega_1\Omega_1} = \text{RBMC}(\vect{Q}_{\Omega_1\Omega_1}, \vect{x}_{S})$
    \State $\qquad\qquad+\Var(\vect{Q}_{\Omega_1\Omega_1}^{-1}\vect{Q}_{\Omega_1S}^{}\vect{x}_{S}^{})$
    \State $\vect{\Sigma}_{\Omega_2\Omega_2} = \text{RBMC}(\vect{Q}_{\Omega_2\Omega_2}, \vect{x}_{S})$
    \State $\qquad\qquad+\Var(\vect{Q}_{\Omega_2\Omega_2}^{-1}\vect{Q}_{\Omega_2S}^{}\vect{x}_{S}^{})$
    \State $\vect{\Sigma}_{SS}^{} = \Var(\vect{x}_{S}^{})$
    \State Return $\vect{\Sigma}$
\end{algorithmic}
\textbf{Output}: $\vect{\Sigma}$
\end{algorithm}

The algorithm computes the diagonal blocks of the covariance matrix using a divide-and-conquer tactics. It can be further simplified to compute only the diagonal entries. The only costly computations here are inverting a matrix $\vect{Q}$ at the bottom of the recursion and evaluating $\vect{Q}_{\Omega_1\Omega_1}^{-1}\vect{Q}_{\Omega_1S}^{}\vect{x}_{S}^{}$ and $\vect{Q}_{\Omega_2\Omega_2}^{-1}\vect{Q}_{\Omega_2S}^{}\vect{x}_{S}^{}$. The base case inversion can be done efficiently using a direct solver, whereas Krylov methods can be used to solve linear systems with $\vect{Q}_{\Omega_1\Omega_1}$ and $\vect{Q}_{\Omega_2\Omega_2}$.

\subsubsection{Parallel RBMC}

Although conceptually simple, the recursive implementation would be inefficient and hard to implement. Setting up a preconditioned Krylov solver for each $\vect{Q}_{\Omega_1\Omega_1}$ at each level would have a huge overhead. In addition, a thorough bookkeeping would be needed for complicated ownership patterns of the original matrix $\vect{Q}$ by different processes. Ideally, each subproblem inherits a processor from the parent problem, but it is unclear how the separators should be distributed.

A better approach would be to flatten out the recursion, and tackle the problem from the domain decomposition point of view. Consider partitioning the graph of $\vect{Q}$ into $n_p$ parts using nested dissection. Denote $\Omega_p$ to be the partitions and $S$ to be a union of vertex separators. The same partitioning is obtained at leaf nodes if we naively apply the recursive RBMC algorithm. Important part is that given $\vect{x}_{S}^{}$, all of $\vect{x}_{\Omega_p}$ are conditionally independent from each other. Then we can rewrite the algorithm in a more parallel friendly way.
\begin{algorithm}[ht]
\caption{Parallel RBMC}\label{alg:var:2}
\textbf{Input}: $\vect{Q}$
\begin{algorithmic}
    \State $\vect{\Sigma} = \vect{0}$
    \State Partition the graph into non-overlapping $\Omega_p$, separated by $S$
    \For{$p=1,\dots,n_p$ \textbf{in parallel}}
        \State $\vect{\Sigma}_{\Omega_p\Omega_p}=\vect{Q}_{\Omega_p\Omega_p}^{-1}+\Var(\vect{Q}_{\Omega_p\Omega_p}^{-1}\vect{Q}_{\Omega_pS}^{}\vect{x}_{S}^{})$
    \EndFor
    \State $\vect{\Sigma}_{SS}^{} = \Var(\vect{x}_{S}^{})$
    \State Return $\vect{\Sigma}$
\end{algorithmic}
\textbf{Output}: $\vect{\Sigma}$
\end{algorithm}

The one crucial difference of this variant from the recursive is that the partitioning of the graph is done only once. This helps to distribute the problem more efficiently and avoid unnecessary movement of data. In addition, each subproblem has exactly one operator $\vect{Q}_{\Omega_p\Omega_p}$, which is factored for the direct solution $\vect{Q}_{\Omega_p\Omega_p}^{-1}$ and can be reused when computing the variance of $\vect{Q}_{\Omega_p\Omega_p}^{-1}\vect{Q}_{\Omega_pS}^{}\vect{x}_{S}^{}$.
\begin{figure}[ht]
    \centering
    \includegraphics[width=0.7\linewidth]{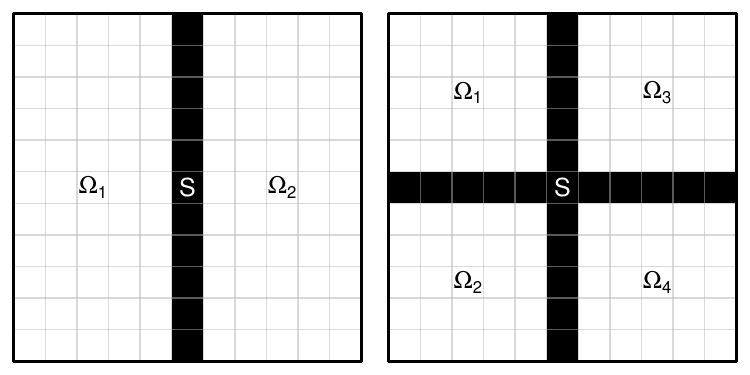}
    \caption{Schur's complement method for 2 and 4 subdomains}
    \label{fig:var:2}
\end{figure}

The algorithm is a stochastic version of the iterative substructuring, also called the Schur complement method. The substructuring method computes $\Var(\vect{x}_{S}^{})$ as an inverse of the Schur complement
\begin{equation}\label{eq:var:4}
    \vect{\Sigma}_{SS}^{} = \Big(\vect{Q}_{SS} -\sum_{p=1}^{n_p}\vect{Q}_{S\Omega_p}\vect{Q}_{\Omega_p\Omega_p}^{-1}\vect{Q}_{\Omega_pS}^{} \Big)^{-1}.
\end{equation}
As we see, the Schur complement itself can be computed and assembled in parallel, by solving with $\vect{Q}_{\Omega_p\Omega_p}$. However, the Schur complement is a dense $|S|\times|S|$ matrix in general, and the dimensions of the separator is $\mathcal{O}(n^{2/3})$ for 3D problems. For large scale problems, this method can quickly become prohibitive. We avoid such bottlenecks, at a cost of accuracy, by approximating the variance using samples from $\vect{x}_{S}^{}$.

\subsubsection{Overlapping RBMC}

Approximating the variance term in Algorithm \ref{alg:var:2} via sampling introduces a Monte Carlo error. The uncertainty in $\vect{x}_{S}^{}$ propagates to the rest of the partition $\Omega_p$ as well. The error can be decreased by increasing the number of samples of $\vect{x}_{S}^{}$. Alternatively, it can also be reduced by pushing the separator outwards, which was also done in \citet{siden2018efficient}. Doing this lessens the dependency of the interior parts of $\Omega_p$ from $S$, since the correlation has to travel further. The similar idea has been used to absorb the effects from the boundary for the SPDE approach on compact domains \citep{lindgren2011explicit}.

The idea can be better explained on a simpler one-dimensional example. Consider a stationary auto-regressive AR(1) process $x_i=\phi x_{i-1}+\epsilon_i$ for $i=1,\dots,99$ with $\phi=0.95$. Let $\vect{x}_{\Omega_1}=(x_1,\dots,x_{49})$, $\vect{x}_{\Omega_2}=(x_{51},\dots,x_{99})$ and $\vect{x}_S=(x_{50})$ be disjoint partitions obtained in Algorithm \ref{alg:var:2}. Now we define two overlapping partitions $\vect{x}_{\Omega_1^*}=(x_{1},\dots,x_{59})$ and $\vect{x}_{\Omega_2^*}=(x_{41},\dots,x_{99})$, created by extending the partitions by $n_l=10$ lags. The corresponding separators are $\vect{x}_{S_1}=(x_{60})$ and $\vect{x}_{S_2}=(x_{40})$. For both partitioning schemes we can compute the RBMC estimates of marginal variances and the relative root mean square errors (RMSE). The RMSE of MC and RBMC estimators are $\sqrt{2/n_k}$ and $\phi^{2n_l}\sqrt{2/n_k}$, for a sample size $n_k$ and a distance $n_l$ from the separator.

\begin{figure*}[ht]
    \centering
    \includegraphics[width=\linewidth]{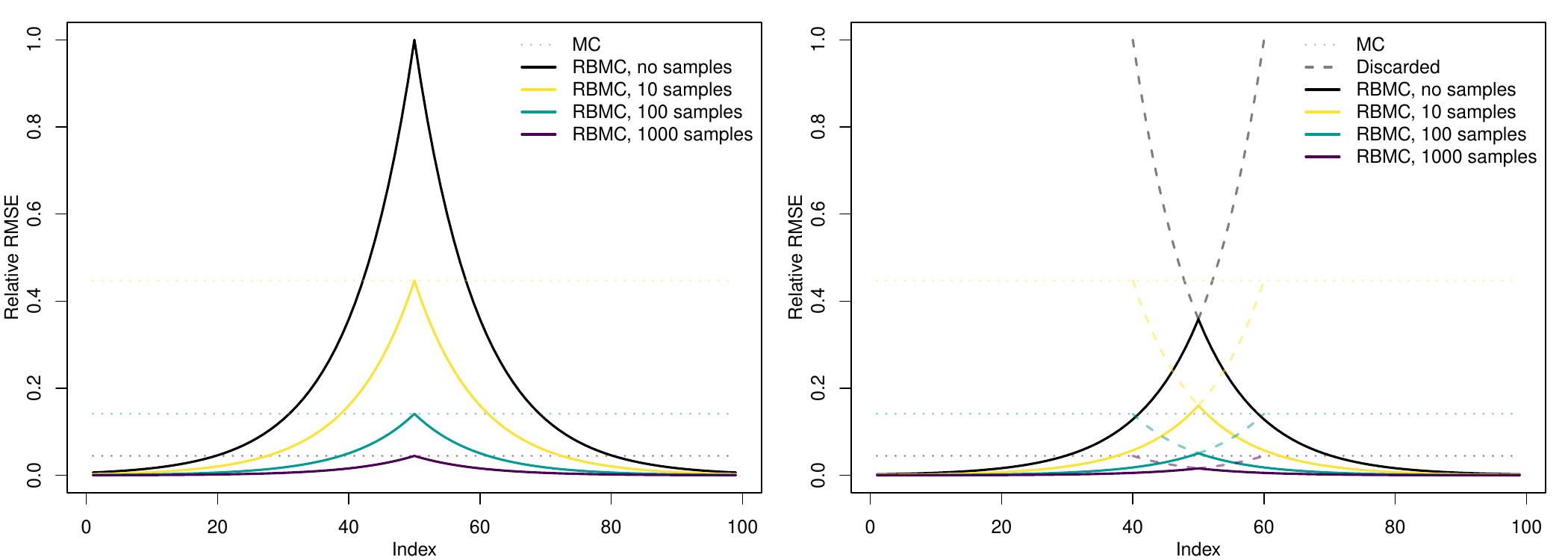}
    \caption{Relative RMSE of parallel (left) and overlapping (right) RBMC estimators with various sample sizes for an AR-1 model with $\phi=0.95$.}
    \label{fig:var:4}
\end{figure*}

\begin{figure}[ht]
    \centering
    \includegraphics[width=0.7\linewidth]{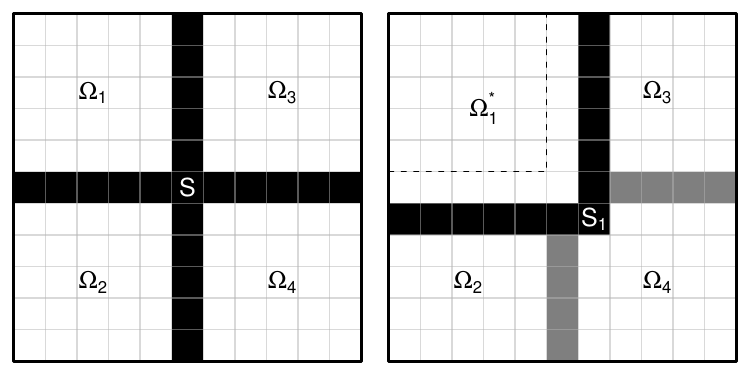}
    \caption{Initial partitioning and the extended subdomain $\Omega_1^*$ ($n_l=1$) with its separating set $S_1$.}
    \label{fig:var:3}
\end{figure}

First, we plot the error for non-overlapping partitions on the left of Figure \ref{fig:var:4}. We can clearly see the error decaying as we move away from the separating set, with the highest error being equal to the MC error exactly at the separator. Then we shift the respective separators by 10 lags in opposite directions and get the second graph in Figure \ref{fig:var:4}. Here we can restrict the variance estimates to match the original partitioning scheme, and discard the values that were on intervals overlapping with the exterior (dashed lines). We will call this estimator the overlapping RBMC. A two-dimensional example is shown in Figure \ref{fig:var:3}.

A good choice of $n_l$ would be the lag, at which correlation drops to zero. For discretely indexed temporal 1D problems it is the temporal correlation $r_t$, for spatial 2D problems on a discrete grid it is the spatial correlation $r_s$. For spatio-temporal 3D models on a continuous domain, it is more complicated. The spatial correlation will not decrease by the same amount as the temporal correlation for the same graph distance. A better strategy would be to have separate overlap parameters $n_{l,s}$ and $n_{l,t}$ based on graph distance at which $r_s\approx0$ and $r_t\approx0$. Although two range parameters interact when the field is non-separable, the actual temporal range is usually smaller than $r_t$. In addition, the posterior range tends to be smaller than the prior, as the data introduces some degree of diagonal dominance to the precision matrix. This leads to unnecessarily larger partitions, but makes the overlap construction somewhat robust, since we always cover the high correlation regions.

Now we present an overlapping version of the RBMC estimator in Algorithm \ref{alg:var:3}. For the sake of simplicity, the algorithm has only one overlap parameter $n_l$. First, denote $N(\Omega_p)$ to be the indices of graph neighbors of $\vect{x}_{\Omega_p}$, i.e. the nodes at an exact distance 1. Then we can generate the overlapping regions $\Omega_p^*$ by extending the original partitions $\Omega_p$ to its neighbor nodes at a distance $n_l$. The separating sets $S_p$ are $(n_l+1)$-th order neighbors, computed in a similar fashion. Then we proceed with computing marginal variances for overlapping partitions $\Omega_p^*$. At the end of the algorithm, we apply a diagonal partition of unity operator $\vect{P}_{p}$, which selects rows of $\Sigma_{\Omega_p^*\Omega_p^*}$ corresponding to indices of non-overlapping $\Omega_p$. The two-dimensional example of such extensions and overlaps is illustrated in Figure \ref{fig:var:3}, where we partition a square region into 4 parts and compute separator for each of them.
\begin{algorithm}[ht]
\caption{Overlapping RBMC}\label{alg:var:3}
\textbf{Input}: $\vect{Q}$
\begin{algorithmic}
    \State Partition the graph into non-overlapping $\Omega_p$
    \For{$p=1,\dots,n_p$ \textbf{in parallel}}
        \State Set $\Omega_p^*=\Omega_p$ and $S_p=N(\Omega_p)\setminus \Omega_p$
        \For{$l=1,\dots,n_l$}
            \State $\Omega_p^*=\Omega_p^*\cup S_p$
            \State $S_p=N(\Omega_p^*)\setminus \Omega_p^*$
        \EndFor
        \State $\vect{\Sigma}_{\Omega_p\Omega_p^*}=\vect{P}_{p}\left(\vect{Q}_{\Omega_p^*\Omega_p^*}^{-1}+\Var(\vect{Q}_{\Omega_p^*\Omega_p^*}^{-1}\vect{Q}_{\Omega_p^*S_p}\vect{x}_{S_p}^{})\right)$
    \EndFor
    \State Return $\vect{\Sigma}$
\end{algorithmic}
\textbf{Output}: $\vect{\Sigma}$
\end{algorithm}

We can get a simplified version of the algorithm for the diagonal elements only, which is given in \ref{alg:var:4}. As we see in Algorithm \ref{alg:var:4}, the main bulk of computations inside an outer loop is done in parallel. The algorithm consists of three main phases: computing the extension and a separator, a direct factorization, and a sampling and correction phase. Graph partitioning can be done efficiently using either the nested dissection algorithm or the multilevel partitioning techniques, both of which are implemented in METIS/ParMETIS library. Computing the neighbor nodes from the adjacency list is very straightforward, and can be done in parallel, if each processor stores the full list. Otherwise, some communication must be done to retrieve such information. For the direct solution $\diag(\vect{Q}_{\Omega_p^*\Omega_p^*}^{-1})$, a highly efficient selected inversion algorithm \citep{verbosio2017enhancing} can be used. Selected inversion scales with the number of nonzeros in $\vect{Q}_{\Omega_p^*\Omega_p^*}$ and has already been implemented in the PARDISO and MUMPS solvers. The factorization from this phase can then be reused for computing the correction in the sampling phase. Since we only need the diagonal of the sample variance, computations may be expressed in terms of elementwise products $\odot$. The samples can be generated using a Krylov solver, which we will discuss soon.
\begin{algorithm}[ht]
\caption{Overlapping RBMC for diagonal elements}\label{alg:var:4}
\textbf{Input}: $\vect{Q}$
\begin{algorithmic}
    \State Partition the graph into non-overlapping $\Omega_p$
    \For{$p=1,\dots,n_p$ \textbf{in parallel}}
        \State Set $\Omega_p^*=\Omega_p$ and $S_p=N(\Omega_p)\setminus \Omega_p$
        \For{$l=1,\dots,n_l$}
            \State $\Omega_p^*=\Omega_p^*\cup S_p$
            \State $S_p=N(\Omega_p^*)\setminus \Omega_p^*$
        \EndFor
        \State $\diag(\vect{\Sigma}_{\Omega_p^*\Omega_p^*}) = \diag(\vect{Q}_{\Omega_p^*\Omega_p^*}^{-1})$
        \For{$k=1,\dots,n_k$}
            \State Sample $\vect{x}^{(k)}$ using a Krylov solver
            \State $\diag(\vect{\Sigma}_{\Omega_p^*\Omega_p^*}) = \diag(\vect{\Sigma}_{\Omega_p^*\Omega_p^*})$
            \State \quad\quad\quad $+\frac{1}{n_k} \left(\vect{Q}_{\Omega_p^*\Omega_p^*}^{-1}\vect{Q}_{\Omega_p^*S_p}\vect{x}_{S_p}^{(k)}\right)$
            \State \quad\quad\quad\quad$\odot \left(\vect{Q}_{\Omega_p^*\Omega_p^*}^{-1}\vect{Q}_{\Omega_p^*S_p}\vect{x}_{S_p}^{(k)}\right)$
        \EndFor
        \State $\diag(\vect{\Sigma}_{\Omega_p\Omega_p}) = \vect{P}_{p}\diag(\vect{\Sigma}_{\Omega_p^*\Omega_p^*})$
    \EndFor
    \State Return $\diag(\vect{\Sigma})$
\end{algorithmic}
\textbf{Output}: $\diag(\vect{\Sigma})$
\end{algorithm}

\subsection{Sampling}

We note that the algorithm requires one to be able to sample from the separating set $\vect{x}_{S}^{}$. In \citet{siden2018efficient}, authors can represent the posterior precision in the form $\vect{Q}=\vect{L}\vect{L}^\tran+\vect{R}\vect{R}^\tran$ for some factors $\vect{L}$ and $\vect{R}$. Then, as suggested by \citet{papandreou2010gaussian}, one can easily sample from the posterior by solving $\vect{Q}\vect{x}=\vect{L}\vect{z}+\vect{R}\vect{w}$ using the conjugate gradient, where $\vect{z}$ and $\vect{w}$ are standard Gaussian samples. However, this cannot be done for diffusion based models as was discussed in Section \ref{sec:2}. Instead, we will use the Lanczos quadrature to evaluate the inverse square root of the precision operator $\vect{Q}^{-1/2}\vect{x}$.

\subsubsection{Lanczos quadrature for matrix functions}

The Lanczos process lays in the heart of Krylov methods for symmetric matrices. The process generates the Krylov subspace $\mathcal{K}_m(\vect{Q},\vect{v}^{(0)})$ by evaluating matrix vector products $\vect{Q}\vect{v}^{(j)}$ and orthogonalizing them against previously computed vectors. So far, we have used the Krylov solvers as a polynomial approximation to the true solution $\vect{Q}^{-1}\vect{z}\approx q(\vect{Q})\vect{z}$. The same can be done for approximating matrix functions $f(\vect{Q})\vect{z}$, by projecting $f(\vect{Q})\vect{z}$ onto the Krylov subspace. The approximation is optimal 2-norm sense \citep{chow2014preconditioned} and is
\begin{equation}\label{eq:sampling:2}
    f(\vect{Q})\vect{z}\approx\vect{V} f(\vect{T})(\eta\vect{e}_1),
\end{equation}
where $\vect{V}$ are orthogonalized Krylov vectors starting with $\vect{v}^{(0)}=\vect{z}/\eta$, $\eta=\|\vect{z}\|$, $\vect{T}$ is a small tridiagonal matrix of Lanczos coefficients and $\vect{e}_1$ is the first column of an $m\times m$ identity matrix. 

The above is also nicely connected to approximating the Riemann-Stieltjes integral representation of the quadratic form
\begin{equation}\label{eq:sampling:3}
\begin{aligned}
    \vect{z}^\tran f(\vect{Q})\vect{z}
    & = \vect{z}^\tran\vect{\Upsilon}f(\vect{\Lambda})\vect{\Upsilon}^\tran\vect{z} \\
    & = \sum_{i=1}^{n} f(\lambda_i) \omega_i^2 \\
    & = \int f(t)d\omega(t),
\end{aligned}
\end{equation}
where $\vect{Q}=\vect{\Upsilon\Lambda\Upsilon}^\tran$ is the eigendecomposition, $w_i=(\vect{\Upsilon}^\tran\vect{z})_i$, and $\omega(t)$ is a piece-wise constant measure representing the distribution of $\omega_i^2/\eta^2$ \citep{golub1994matrices}\citep{ubaru2017fast}. Then, the link can be established by approximating the integral with Gauss quadrature
\begin{equation}\label{eq:sampling:4}
    \int f(t)d\omega(t) \approx \sum_{j=1}^m (\eta\Tilde{\upsilon}_{1j})^2 f(\Tilde{\lambda}_j),
\end{equation}
where the nodes and weights can be elegantly obtained from the Lanczos process \citep{golub1994matrices}. Here, $\Tilde{\lambda}_j$ is the $j$-th eigenvalue of the Lanczos' tridiagonal matrix $\vect{T}$, and $\Tilde{\upsilon}_{1j}$ is the first entry of the corresponding eigenvector. By denoting $\vect{T}=\Tilde{\vect{\Upsilon}}\Tilde{\vect{\Lambda}}\Tilde{\vect{\Upsilon}}^\tran$, we can reassemble the approximation above as
\begin{equation}\label{eq:sampling:5}
\begin{aligned}
    \vect{z}^\tran f(\vect{Q})\vect{z}
    & \approx \sum_{j=1}^m (\eta\Tilde{\upsilon}_{1j})^2 f(\Tilde{\lambda}_j) \\
    & = (\eta\vect{e}_1)^\tran \Tilde{\vect{\Upsilon}}f(\Tilde{\vect{\Lambda}})\Tilde{\vect{\Upsilon}}^\tran (\eta\vect{e}_1) \\
    & = (\eta\vect{e}_1)^\tran f(\Tilde{\vect{\Upsilon}}\Tilde{\vect{\Lambda}}\Tilde{\vect{\Upsilon}}^\tran) (\eta\vect{e}_1) \\
    & = \vect{z}^\tran\vect{V} f(\vect{T})(\eta\vect{e}_1),
\end{aligned}
\end{equation}
which we can recognize as a product between $\vect{z}^\tran$ and the approximation to $f(\vect{Q})\vect{z}$.

\subsubsection{Preconditioned Krylov sampling}

We could use the formula in Equation \eqref{eq:sampling:2} to sample from a posterior GMRF by taking $\vect{Q}$ as our operator, $\vect{z}$ as the right hand side, and $f(z)=z^{-1/2}$ as our function. As long as the dimension of the Krylov subspace stays low, it is rather easy to compute $f(\vect{T})=\vect{T}^{-1/2}$ through the eigendecomposition to find weights for $\vect{V}$. On the contrary, if the condition number of $\vect{Q}$ is large, it results in longer iterations and higher memory requirements.  Therefore, preconditioning is a crucial step to keep the algorithm efficient. However, the sampling formula in Equation \eqref{eq:sampling:2} becomes more complicated, and will depend on the type of preconditioner we will use. We will further follow the results of \citet{chow2014preconditioned}. 

We consider a symmetric split preconditioning obtained from incomplete factorization of $\vect{Q}$. For example, an incomplete Cholesky with zero fill-in IC(0) \citep{saad2003iterative}, or its block version called block Jacobi. The preconditioned operator then becomes $\Tilde{\vect{Q}}=\vect{L}^{-1}\vect{Q}\vect{L}^{-\tran}$, and we can evaluate an inverse square root buy building the subspace $\mathcal{K}_m(\Tilde{\vect{Q}},\vect{z})$,
\begin{equation}\label{eq:sampling:6}
    \Tilde{\vect{x}} = \Tilde{\vect{Q}}^{-1/2}\vect{z} \approx \vect{V}\vect{T}^{-1/2}(\eta\vect{e}_1).
\end{equation}
The vector $\Tilde{\vect{x}}$ has an approximate covariance
\begin{equation}\label{eq:sampling:7}
    \Tilde{\vect{Q}}^{-1} = (\vect{L}^{-1}\vect{Q}\vect{L}^{-\tran})^{-1} = \vect{L}^\tran\vect{Q}^{-1}\vect{L},
\end{equation}
that is, $\Tilde{\vect{x}}\sim\mathcal{N}(\vect{0},\vect{L}^\tran\vect{Q}^{-1}\vect{L})$. To rewind the effect of the preconditioner, we only need to be able to apply $\vect{L}^{-\tran}$. Then
\begin{equation}\label{eq:sampling:8}
    \vect{x}
    = \vect{L}^{-\tran}\Tilde{\vect{x}} 
    \approx \vect{L}^{-\tran}\vect{V}\vect{T}^{-1/2}(\eta\vect{e}_1)
\end{equation}
has an approximately Gaussian density $\vect{x}\sim\mathcal{N}(\vect{0},\vect{Q}^{-1})$. The Algorithm \ref{alg:sampling:1} below generates $K$ samples for $\vect{x}$, from which we can select the desired elements $\vect{x}_{S_p}^{}$ for our overlapping RBMC estimator in Algorithm \ref{alg:var:3}.
\begin{algorithm}[ht]
\caption{Preconditioned Krylov sampler}\label{alg:sampling:1}
\textbf{Input}: $\vect{Q}$
\begin{algorithmic}
    \State Compute a factored symmetric preconditioner $\vect{LL}^\tran\approx\vect{Q}$
    \For{$k=1,\dots,n_k$}
        \State Generate $\vect{z}^{(k)}\sim\mathcal{N}(\vect{0},\vect{I})$
        \State Construct $\mathcal{K}(\Tilde{\vect{Q}},\vect{z}^{(k)})$ and obtain $\vect{T}$, $\vect{V}$ using preconditioned MINRES/GMRES
        \State Eigendecomposition $\vect{T}=\Tilde{\vect{U}}\Tilde{\vect{\Lambda}}\Tilde{\vect{U}}^\tran$
        \State Compute $\Tilde{\vect{x}}=\vect{V}\Tilde{\vect{\Upsilon}}\Tilde{\vect{\Lambda}}^{-1/2}\Tilde{\vect{\Upsilon}}^\tran(\eta\vect{e}_1)$
        \State Rewind the preconditioner $\vect{x}^{(k)}=\vect{L}^{-\tran}\Tilde{\vect{x}}$
    \EndFor
\end{algorithmic}
\textbf{Output}: $\{\vect{u}_1,\dots,\vect{u}_{n_k}\}$
\end{algorithm}

Several practical consideration must be taken into account when implementing the sampler. First, we must solve a system for each new $\vect{z}^{(k)}$. Reusing the same subspace for different samples will produce lower quality samples. Such an approach corresponds to sampling from a low-rank approximation $\Tilde{\vect{Q}}^{-1}\approx\vect{V}\vect{T}^{-1}\vect{V}^\tran$, which is very innacurate unless $m$ is close to $n$. For multiple right hand sides, we must generate the subspace and perform eigendecomposition each time we solve a system. Although the cost is higher, the quality of samples will be much better.

Second, the preconditioned right hand side must be $\vect{z}^{(k)}$, meaning that the original unpreconditioned problem we must solve is $\vect{Q}\vect{x}=\vect{L}\vect{z}^{(k)}$. Therefore, when using Krylov solvers, one must premultiply the right hand side by an incomplete factor $\vect{L}$ before an actual solve. However, Krylov solvers (such as PETSc) provide implementations only for applying $\vect{L}^{-1}$ or $\vect{L}^{-\tran}$, but never $\vect{L}$. In such cases, one might need to customize the solver or the preconditioner, to pass $\vect{z}$ directly as a preconditioned right hand side.

Third, the eigendecomposition of $\vect{T}^{-1/2}$ cannot be replaced by a solve with a Cholesky factor. Although both options are valid inverse square root operators, our approximation implicitly relies on quadrature nodes and weights obtained from the eigendecomposition of $\vect{T}$. In practice, the decomposition is performed by each processor instead of parallelizing it. This saves communication cost and the computational overhead is not big. The structure of $\vect{T}$ becomes very convenient, since we do not have to tridiagonalize the matrix $\vect{T}$ for an eigendecomposition, and instead we can call optimized LAPACK routines directly.

Fourth, there are no natural tools to measure closeness of matrix function approximations when using Krylov solvers. Some analysis has been done for decaying matrix functions \citep{frommer2021analysis}. However, in practice, the residuals of a solution $\vect{Q}^{-1}\vect{z}$ are used as a criterion, despite $f(\vect{Q})\vect{z}$ converging differently in general.

\subsection{Hyperparameter estimation}

Now assume the hyperparameters are unknown and we want the mode of the marginal density $\pi(\vect{\theta}|\vect{y})$. First, we impose a prior $\pi(\vect{\theta})$ on hyperparameters of the LGM model. One can use penalizing complexity (PC) priors \cite{simpson2017penalising} on the precision of the likelihood $\tau_y$, log-range $\log r_s$, $\log r_t$ and log-variance $\sigma^2$ of the spatio-temporal field (see \cite{lindgren2020diffusion}). Then the marginals of the hyperparameters can be computed as a ratio of the joint and the full conditional, which is exact in Gaussian likelihood case and is called Laplace approximation for other likelihoods \cite{rue2009approximate, rue2017bayesian}.
\begin{equation}\label{eq:hyperpar:1}
\begin{aligned}
    \pi(\vect{\theta}|\vect{y}) \propto \frac{\pi(\vect{y}|\vect{x},\vect{\theta})\pi(\vect{x}|\vect{\theta})\pi(\vect{\theta})}{\pi(\vect{x}|\vect{y},\vect{\theta})}\bigg|_{\vect{x}=\vect{\mu}(\vect{\theta})}
\end{aligned}
\end{equation}

\subsubsection{Numerical optimization}

What we have done so far, is to approximate the mean $\vect{\mu}(\vect{\theta})$ and selected entries of the variance $(\vect{Q}(\vect{\theta}))^{-1}$ under fixed hyperparameter configuration $\vect{\theta}$. However, if we want to find the mode of the marginal $\pi(\vect{\theta}|\vect{y})$, we have to use a gradient based optimization algorithm. The gradient ascent/descent and the Newton method are two common choices, and can be written using a single formula
\begin{equation}\label{eq:hyperpar:2}
    \vect{\theta}^{(k+1)} = \vect{\theta}^{(k)} - \rho^{(k)}\cdot(\vect{H}^{(k)})^{-1}\nabla\log\pi(\vect{\theta}^{(k)}|\vect{y}).
\end{equation}
where $\rho^{(k)}$ is a learning rate parameter. For general references, please see \cite{amari1993backpropagation, kiwiel2001convergence, bottou1998online}. Choosing the preconditioner to be the negative identity $\vect{H}^{(k)}=-\vect{I}$ yields the gradient descent/ascent algorithm, the exact Hessian $\vect{H}^{(k)}=\nabla^2\log\pi(\vect{\theta}^{(k)}|\vect{y})$ results in (damped) Newton method and an approximation to the Hessian gives us so-called quasi-Newton methods. When the gradient is randomized the algorithm is essentially a stochastic gradient descent or a stochastic Newton method. For the guaranteed convergence of the stochastic gradient descent, $\rho^{(k)}$ has to to diminish with $k$ and the stochastic gradient must have bounded variance. Also, the stochastic Newton can be unstable due to noise and is biased unless we have the exact full Hessian, which is not possible in our context. However, one could employ an approximate but deterministic preconditioner $\vect{H}^{(k)}$ that would still be better than uniform scaling of the gradient descent. We use a diagonal matrix of second derivatives, obtained from two first derivatives with stochastic components removed.

\subsubsection{Gradient estimation}

Next, we show how we can approximate the gradient using the RBMC estimator of the partial inverse. Three densities in the Equation \eqref{eq:hyperpar:1} except $\pi(\vect{\theta})$ are Gaussian. Thus, it will involve computing the quantities $\dtheta{j}\log|\vect{Q}_y(\vect{\theta})|$, $\dtheta{j}\log|\vect{Q}_x(\vect{\theta})|$, $\dtheta{j}\log|\vect{Q}(\vect{\theta})|$, as well as $\dtheta{j}(\vect{y}-\vect{Ax})^\tran\vect{Q}_y(\vect{\theta})(\vect{y}-\vect{Ax})$, $\dtheta{j}\vect{x}^\tran\vect{Q}_x(\vect{\theta})\vect{x}$, $\dtheta{j}(\vect{x}-\vect{\mu}(\vect{\theta}))^\tran\vect{Q}(\vect{\theta})(\vect{x}-\vect{\mu}(\vect{\theta}))$ for the partial derivatives. The quadratic forms can be computed directly using the finite differences, but the log-determinants are harder to approximate. We use the following fact
\begin{equation}\label{eq:hyperpar:3}
\begin{aligned}
    \dtheta{j}\log|\vect{Q}(\vect{\theta})| 
    &= \dtheta{j}(\log|\vect{Q}_{uu}(\vect{\theta})|+\log|\vect{S}(\vect{\theta})|) \\
    &= \tr(\vect{Q}_{uu}(\vect{\theta})^{-1}\dtheta{j}\vect{Q}_{uu}(\vect{\theta})) \\
    &\qquad+ \tr(\vect{S}(\vect{\theta})^{-1}\dtheta{j}\vect{S}(\vect{\theta}))
\end{aligned}
\end{equation}
where $\vect{S}(\vect{\theta})$ is the Schur complement of $\vect{Q}_{uu}(\vect{\theta})$ in $\vect{Q}(\vect{\theta})$. The trace of the latter can be computed using dense linear algebra, since the dimensions of the matrix $(n_{\beta}\times n_{\beta})$ are assumed to be small. On the other hand, the first term can be expressed as sum over the elementwise product of the inverse and the partial derivative, meaning that we have to know the inverse $\vect{Q}_{uu}(\vect{\theta})^{-1}$ only at nonzero positions of $\dtheta{j}\vect{Q}_{uu}(\vect{\theta})$ or $\vect{Q}_{uu}(\vect{\theta})$, i.e. the selected inverse. We can compute the trace in parallel, because the precision matrix, its inverse and derivatives are already distributed along corresponding rows $\Omega_p$.
\begin{equation}\label{eq:hyperpar:4}
\begin{aligned}
    & \tr(\vect{Q}_{uu}(\vect{\theta})^{-1}\dtheta{j}\vect{Q}_{uu}(\vect{\theta})) = \\
    &\qquad= \sum_{i,k}^{} \left[\vect{Q}_{uu}(\vect{\theta})^{-1}\odot\dtheta{j}\vect{Q}_{uu}(\vect{\theta})\right]_{ik} \\
    &\qquad= \sum_{p=1}^{n_p}\sum_{i,k}^{} \bigg[\Big(\vect{Q}_{uu}(\vect{\theta})^{-1}\Big)_{\Omega_p\Omega_p^*} \\
    &\qquad\qquad\qquad \odot\Big(\dtheta{j}\vect{Q}_{uu}(\vect{\theta})\Big)_{\Omega_p\Omega_p^*}\bigg]_{ik} 
\end{aligned}
\end{equation}
Therefore, we can use our selected inversion method in Algorithm \ref{alg:var:3} to compute partial derivatives of $\log|\vect{Q}(\vect{\theta})|$ and $\log|\vect{Q}_x(\vect{\theta})|$, and hence, of $\log\pi(\vect{\theta}|\vect{y})$. Please find the rough algorithm below (trivial partial derivatives are omitted) in Algorithm \ref{alg:hyperpar:1}.
\begin{algorithm}[ht]
\caption{Hyperparameter estimation}\label{alg:hyperpar:1}
\textbf{Input}: $\vect{\theta}^{(0)}$
\begin{algorithmic}
    \For{$k=1,\dots,\text{convergence}$}
        \State Solve for $\vect{\mu}(\vect{\theta})$ and $\vect{S}(\vect{\theta})$
        \State $\vect{Q}(\vect{\theta})^{-1} \approx \text{RBMC}(\vect{Q}(\vect{\theta}))$
        \State $\vect{Q}_x(\vect{\theta})^{-1} \approx \text{RBMC}(\vect{Q}_x(\vect{\theta}))$
        \For{$j=1,\dots,n_{\theta}$}
            \State Compute $\dtheta{j}\log|\vect{Q}(\vect{\theta})|$, $\dtheta{j}\log|\vect{Q}_x(\vect{\theta})|$ 
            \State \qquad and other terms in the derivative
            \State Compute $\dtheta{j}\log\pi(\vect{\theta}|\vect{y})$
        \EndFor
        \State Update $\rho^{(k)}$, $\vect{H}^{(k)}$
        \State $\vect{\theta}^{(k+1)} = \vect{\theta}^{(k)} - \rho^{(k)}\cdot(\vect{H}^{(k)})^{-1}\nabla\log\pi(\vect{\theta}^{(k)}|\vect{y})$
    \EndFor
    \State Return $\vect{\theta}^{(k)}$
\end{algorithmic}
\textbf{Output}: $\diag(\vect{\Sigma})$
\end{algorithm}

\section{Numerical results}\label{sec:4}

In this section, we will apply the methods described in Section \ref{sec:3} on a simulated data and a real world example. For demonstration purposes, we will consider thin 3D models with small temporal separators, usually found in applications for daily weather data. We will model the spatio-temporal component in both examples by a critical diffusion model with parameters $(\alpha_t,\alpha_s,\alpha)=(1,2,1)$ \citep{lindgren2020diffusion}. In the following examples, we consider both the fixed and unknown hyperparameters cases, in latter setting, we impose the PC prior on $\vect{\theta}$.
\begin{equation}\label{eq:num:1}
\begin{aligned}
    \vect{y}|\vect{x},\vect{\theta} &\sim \mathcal{N}(\vect{Ax},\vect{Q}_{y}^{-1})\\
    \vect{x}|\vect{\theta} &\sim \mathcal{N}(\vect{0},\vect{Q}_{x}^{-1})\\
    \vect{\theta} &\sim \pi_{\text{PC}}(\vect{\theta})
\end{aligned}
\end{equation}
The full conditional read as
\begin{equation}\label{eq:num:2}
    \vect{x}|\vect{y},\vect{\theta} \sim \mathcal{N}(\vect{\mu},\vect{Q}^{-1}),
\end{equation}
where $\vect{Q}=\vect{Q}_{x}+\vect{A}^\tran\vect{Q}_{y}\vect{A}$ and $\vect{\mu}=\vect{Q}^{-1}\vect{A}^\tran\vect{Q}_{y}\vect{y}$ depend on $\vect{\theta}$. The posterior mean and marginal variances are computed according to the Equation \eqref{eq:mean:5} and Algorithms \ref{alg:var:4} and \ref{alg:sampling:1}. The implementation of the algorithms was written in C using the PETSc library \citep{petsc-efficient} and MUMPS as the direct solver \citep{amestoy2001fully}. While PARDISO \cite{schenk2001pardiso} is superior, we use MUMPS because it is free and open source.

It is important to note that the posterior precision matrix is not assembled explicitly, only the first block $\vect{Q}_{uu}$ corresponding to the spatio-temporal field. This is done to ease the implementation and reduce the load imbalance. The spatio-temporal block itself is computed from the sum of Kronecker products of temporal $\vect{J}$ and spatial $\vect{K}$ matrices. The mesh and the matrices can be generated by the R-INLA package in R language. 

For space-time models that are long in time direction, the partitioning naturally divides the graph into time intervals. In this case, some simplifications can be done to assemble the precision matrices. For instance, if we distribute the rows of temporal $\vect{J}$ matrices to different processes, Kronecker products $\vect{J}\otimes\vect{K}$ can be computed locally as long as $\vect{K}$ is available at each process. This helps to efficiently assemble the matrix in parallel without computing the full Kronecker product. The same goes for a projection matrix $\vect{A}_{u}$, which can be expressed as a Kronecker product between temporal and spatial projections $\vect{A}_{t}\otimes\vect{A}_{s}$.

\subsection{Simulated example}

\begin{figure*}[ht]
    \centering
    \includegraphics[width=0.6\linewidth]{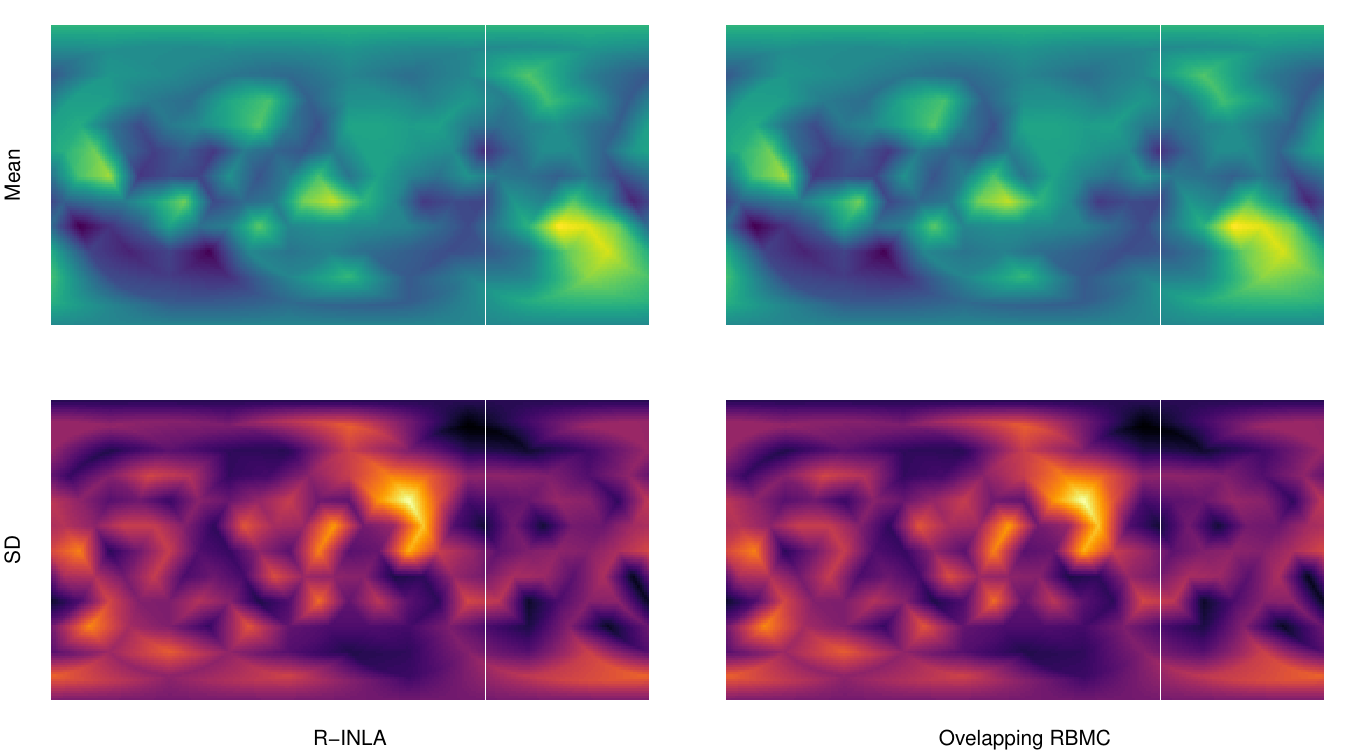}
    \includegraphics[width=0.066\linewidth]{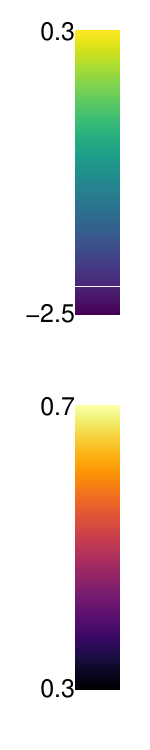}
    \includegraphics[width=0.3\linewidth]{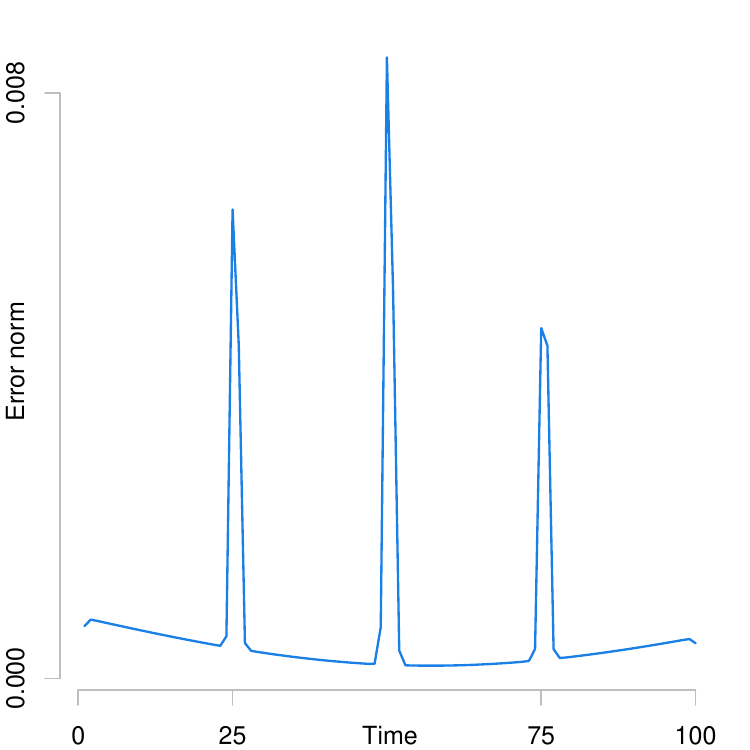}
    \caption{Left: Posterior mean (first row) and standard deviation (second row) at $t=25$ using R-INLA (3.94 sec/fn-call) and Overlapping RBMC (1.47 sec/fn-call). Right: The relative error of overlapping RMBC estimator for the posterior marginal variance versus time.}
    \label{fig:sim:1}
\end{figure*}

Now we present some results for a simulated example, where we generate a synthetic data and fit a space-time model. The observations are generated as $y(\vect{s},t)=1+t+\sin(6\pi t)+\epsilon$, with $1$ and $t$ being the covarites and $\epsilon\sim\mathcal{N}(0,1.01)$. The locations $\vect{s}$ are sampled uniformly on a sphere. The mesh consists of $n_s=162$ nodes on a sphere and $n_t=100$ nodes in time, $n_u=16{,}200$ in total. First, we assume that hyperparameters are fixed to: $r_s=1$, $r_t=10$, $\sigma^2=1$, $\tau_y=\exp(1)$. We divide the field into 4 temporal partitions $\Omega_p$, which are then extended to overlapping $\Omega_p^*$ using $n_l=1$ neighbors. Each partition owns all 162 nodes in space, 25 non-overlapping nodes in time and 26-27 with the overlap, totaling at least 8{,}262 nodes per process. The number of samples for the RBMC estimator is $n_k=5$. We compute the posterior marginals for fixed hyperparameters using R-INLA package and our own approximation from Section \ref{sec:3}. The posterior marginal mean and standard deviation for the field at time $t=25$ (on the boundary of the partition) are visualized in Figure \ref{fig:sim:1}.

As we see, the posterior plots are almost indistinguishable. This is expected due to good conditioning of the precision matrix, despite the slice being near the separating set at $t=26$. To see the overall behavior of the RMBC estimator, let us fix the spatial location and plot the error against time. The maximum relative error of marginal variance for a single spatial node is given on the right in Figure \ref{fig:sim:1}. We clearly see how error rises, as we approach the boundaries of a partition, behaving as described in Section \ref{sec:3}. However, the approximation is also very precise near the separator, and the maximum relative error does not exceed $10^{-2}$. Setting the overlap parameter to be equal to the temporal range $n_l=r_t=10$ would give the error of the magnitude $10^{-12}$. At this distance, the correlation drops to negligible levels and almost no correction is needed from sampling. Of course, one could decrease $n_l$ and increase the number of samples $K$, and still achieve the same level of accuracy. In the end, the parameter choice will depend on not only the temporal range, but also on other hyperparameters, as well as the problem size and hardware constraints.

Now, we relax our assumptions and treat the hyperparameters as unknowns. We optimize for the hyperparameters using Algorithm \ref{alg:hyperpar:1} and our approximation to the gradient. As a reference, we first run the R-INLA package, which produces $\vect{\theta}_{\text{INLA}}^*=(-1.141, 0.503, 0.340, 4.580)$ in 158.2 seconds with 0.347 seconds per function call. In comparison, our optimizer converges to $\vect{\theta}_{\text{RBMC}}^*=(-1.144, 0.501, 0.341, 4.575)$ in 15 iterations and 286.7 seconds (about 1.713 seconds per function call), with relative norm of the gradient $<10^{-3}$ and $\|\vect{\theta}_{\text{RBMC}}^*-\vect{\theta}_{\text{INLA}}^*\|=0.00579$. We used a ``warm'' start $\vect{\theta}_{\text{RBMC}}^{(0)}=(-1,0,0,3)$, a learning rate $\rho=0.9$, a diagonal Hessian approximation from the deterministic part of the RBMC estimator. R-INLA is a highly optimized library and our code is slower for this problem, but it requires much less storage (peak 43 MB opposed to 4.6 GB) and is able to run in a distributed setting.

\subsubsection{Strong and weak scaling}

We now also present running times for the same simulation problem with different number of processes and temporal dimensions. All runs were conducted on Intel nodes of the Ibex cluster at KAUST. The Table \ref{tab:sim:1} shows results for a fixed problem size, with $n_s=162$ and $n_t=1{,}658{,}880$, but differing number of processors (partitions). In an ideal situation, the inference time decreases twice every time we double the number of processors. We see that this is not entirely the case, and with $1024$ processors (on $67$ nodes) we even start to see an increase. At this point, the communication costs start to outweigh the benefits of parallelization.
\begin{table*}[ht]
    \centering
    \caption{Running time for fixed problem size (1,658,880 unknowns)}
    \small
    \begin{tabular*}{\linewidth}{@{\extracolsep\fill}rllllllllll}
        \toprule
        $n_t$ & 10240 & 10240 & 10240 & 10240 & 10240 & 10240 & 10240 & 10240 & 10240 & 10240 \\
        $n_p$ & 2 & 4 & 8 & 16 & 32 & 64 & 128 & 256 & 512 & 1024 \\
        \midrule
        times (s) & 588.62 & 331.84 & 194.09 & 89.06 & 51.12 & 29.53 & 19.73 & 13.40 & 11.16 & 13.30 \\
        \bottomrule
    \end{tabular*}
    
    \label{tab:sim:1}
\end{table*}

\begin{figure*}[ht]
    \centering
    \includegraphics[width=0.9\linewidth]{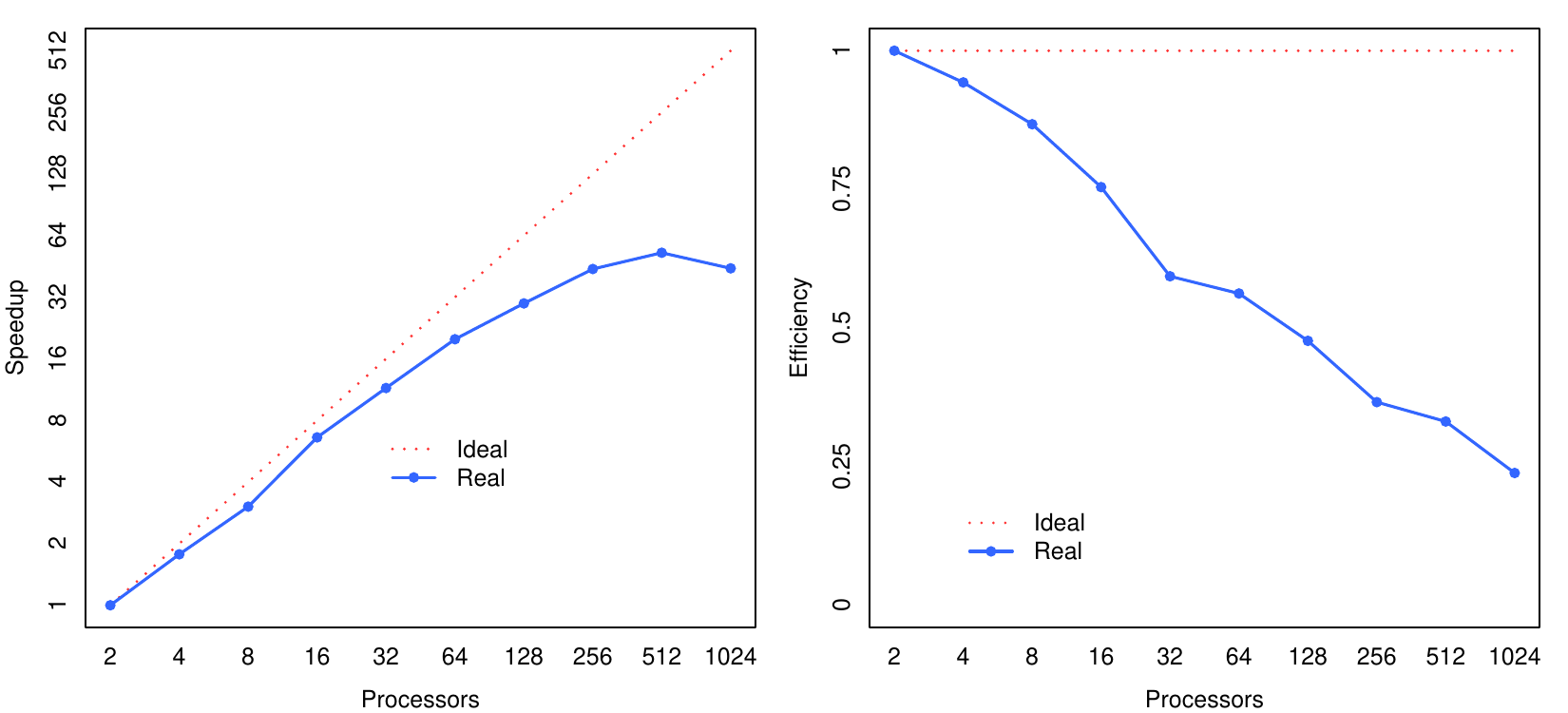}
    \caption{Strong and weak scaling for inference based on overlapping RBMC algorithm}
    \label{fig:sim:3}
\end{figure*}

What we are also interested in is how well the inference scales with fixed problem size per process. In this case, we can measure the overhead and inefficiencies in communication. The Table \ref{tab:sim:2} summarizes the times for various number of processes (partitions). The spatial dimensions of the problem are the same, $n_s=162$. The temporal dimensions per process are fixed to $n_t/n_p=100$, and to $n_t/n_p+2n_l=120$ with the overlap, totalling $19{,}440$ unknowns per process. In total, we have $16{,}588{,}800$ unknown variables. We see that the running time slowly increases, doubling for $128$ cores ($26$ nodes) and quadrupling for $1024$ cores ($75$ nodes). The corresponding plot in Figure \ref{fig:sim:3} shows this steady decrease in efficiency.
\begin{table*}[ht]
    \centering
    \caption{Running time for fixed workload per process 
    (19,440 unknowns)}
    \small
    \begin{tabular*}{\linewidth}{@{\extracolsep\fill}rllllllllll}
        \toprule
        $n_t$ & 200 & 400 & 800 & 1600 & 3200 & 6400 & 12800 & 25600 & 51200 & 102400 \\
        $n_p$ & 2 & 4 & 8 & 16 & 32 & 64 & 128 & 256 & 512 & 1024 \\
        \midrule
        times (s) & 11.71 & 12.42 & 13.50 & 15.53 & 19.74 & 20.83 & 24.56 & 31.97 & 35.34 & 49.10 \\
        \bottomrule
    \end{tabular*}
    \label{tab:sim:2}
\end{table*}

\subsection{Application to US daily temperature data}

We conclude the section by demonstrating the results obtained using our method for the US temperature data \citep{menne2012overview}. The data consists of 12 years (4383 days) of daily temperature observations, recorded at 4730 stations across the continental US. Apart from the time and location information, we are also provided with elevation. We additionally include the y-coordinates and the first harmonics (sine and cosine) as the covariates for seasonal temperature variation. In total, there are 5 covariates including the intercept. The spatio-temporal field is modeled using the critical diffusion (121) model. First, we fit a subset of the original dataset using INLA and our method. The hyperparameters are estimated on a 1-year dataset using 100 stations and mesh sizes $n_s=119$ and $n_t=365$. The hyperparameter vector estimated by R-INLA is $\vect{\theta}_{\text{INLA}}^*=(7.850, 4.529, 2.050, -2.039)$, that is, $r_s=2566.4$, $r_t=92.65$, $\sigma^2=7.768$ and $\tau_y=0.130$. The corresponding RBMC estimate is $\vect{\theta}_{\text{RBMC}}^*=(7.043, 2.867, 1.627, -1.999)$, which is $r_s=1144.8$, $r_t=17.58$, $\sigma^2=5.088$ and $\tau_y=0.135$. We used $\rho=0.5$, $n_k=20$, a diagonal Hessian and ``warm'' starting values $\vect{\theta}_{\text{RBMC}}^{(0)}=(7,0,0,0)$. Larger models could not be run using R-INLA on a computer with only 16 GB of RAM. It took R-INLA 288.9 seconds with peak memory usage of 7.81 GB on 12 threads. On the other hand, our implementation required only 108.7 MB of memory, but it ran for about 1070.3 seconds using 6 cores (6 threads). The run time difference can be due to R-INLA using a deterministic Newton method with line search and being less wasteful, whereas our approach is stochastic and typically needs longer runs.  We also note that the gradient norm of the objective function did not change dramatically in the vicinities of $\vect{\theta}_{\text{INLA}}^*$ and $\vect{\theta}_{\text{RBMC}}^*$, meaning that if we use the gradient norm as a stopping criterion, we might end up in different parts of the target when starting from different initial positions. E.g. we obtain $\vect{\theta}_{\text{RBMC}}^*\approx(5,3,2-2)$ when starting from $(0,0,0,0)$. This might explain the discrepancy in $\vect{\theta}$ estimates.

Now we do the inference on the full dataset using the spatial and temporal mesh generated by R-INLA, which have respective dimensions $n_s=926$ and $n_t=4383$. The total size of the spatio-temporal field is therefore $n=4{,}058{,}658$. Once more, the graph can be easily partitioned along the temporal axis. The overlap parameter is $n_l=1$, the number of partitions (processes) is $n_p=100$. Each process gets about $22$ non-overlapping time points and about $24$ with the overlap. Then the median workload per process is $24\times926=20{,}372$. We set the number of samples to $n_k=10$. The code took $4744.05$ seconds to run on $100$ cores on multiple nodes of KAUST's Ibex cluster. We use the learning rate of $\rho^{(k)}=0.8^{k\bmod10}$, $n_k=10$, a diagonal Hessian and ``cold'' initial values $\vect{\theta}_{\text{RBMC}}^{(0)}=(0,0,0,0)$. The code run until the relative norm of the gradient reached $10^{-3}$. The estimated hyperparameter values are $\vect{\theta}_{\text{RBMC}}^*=(4.492, 0.978, 3.080, -2.324)$.

Below we plot posterior marginals for several slices of the spatio-temporal field at $t\in\{21,22,365,4383\}$, where $t=22$ corresponds to the boundary of the first partition. The posterior mean seems to capture the effects of geographical features on the temperature well. For instance, the western part of the US remains the same throughout both 2-day windows in January and July. At the same time, the central part experiences more rapid changes. The January (left) plots shows the cold front (dark) moving eastwards. The second row shows the posterior marginal standard deviations. The overall picture is not as dynamic, but we can see that the uncertainty is lower in regions with more observations. The variance increases outside of the boundaries, which is very typical for the SPDE approach.
\begin{figure*}[ht]
    \centering
    \includegraphics[width=0.9\linewidth]{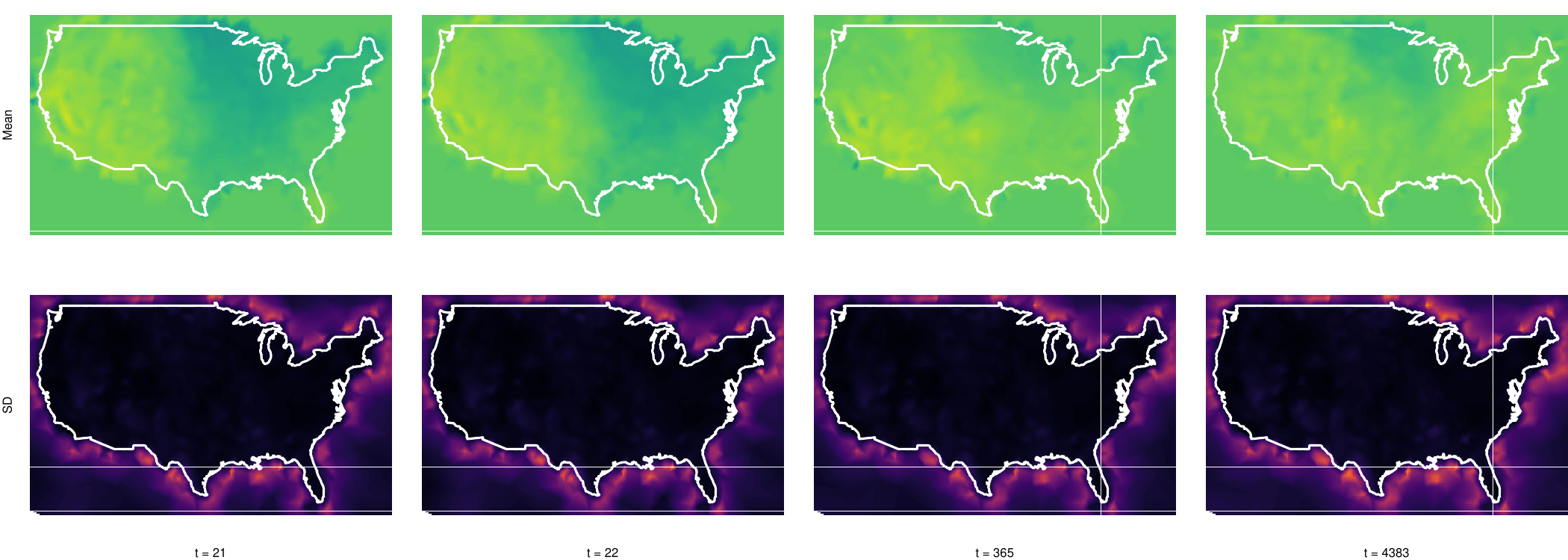}
    \includegraphics[width=0.065\linewidth]{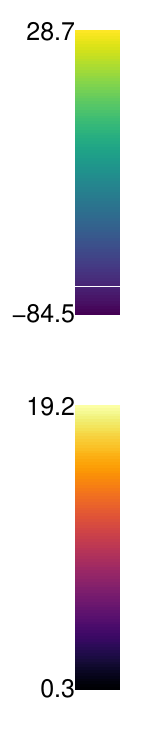}
    \caption{Posterior mean (first row) and standard deviation (second row) at $t\in\{21,22,365,4383\}$ from the overlapping RBMC estimator.}
    \label{fig:app:1}
\end{figure*}

\section{Conclusion}\label{sec:5}

In this paper, we have presented a scalable distributed approach for approximating the posterior marginals of spatio-temporal GMRFs and approximating the hyperparemeters. The proposed methods rely on Krylov solvers for computing the posterior mean and sampling, and a hybrid Rao-Blackwellized Monte Carlo approach for estimating the posterior marginal variances. We then reuse the obtained partial inverse to construct a stochastic gradient direction during the optimization step. The major emphasis was put on the scalability of algorithms in a setting, where the matrices are prohibitively large and are distributed among many cluster nodes. The numerical results demonstrated that the approximation can recover the posterior parameters from R-INLA and the implementation can scale on up to $1024$ cores on $76$ nodes.

Profiling our code also showed that the main computational bottlenecks are direct factorization and sampling. Thus, future optimizations to reduce the total run time may include a more efficient extension of partitions and a more localized strategy for generating samples. For instance, the overlap parameter could be computed dynamically by using probing techniques to approximate the posterior correlation range. Then we can generate more compact overlaps and waste less resources on uncorrelated halo regions. To reduce the communication time during sampling stage, one could try to form samples from the separators directly, rather than solving for global samples. Lastly, we also note that the condition of the precision matrix $\vect{Q}$ is highly dependent on the hyperparameter configuration, which affects overall convergence of Krylov methods, correlation of the field components, and hence the efficiency of any method based on partitioning the graph.

The results can be extended in three main ways. First, the proposed methods considers only Gaussian distributed observations, for which we have closed form posterior densities. Other likelihoods could be added by iterating the Gaussian approximation several times, which corresponds to applying a Newton-Raphson method for non-linear problems. Computationally, this only adds a few more linear solves during iterations. Second, the hyperparameters of the field and the likelihood are approximated only at the mode in our examples. To compute the full posterior distribution of hyperparameters, one can employ a numerical integration scheme as in the INLA methodology \citep{rue2009approximate, rue2017bayesian}. Then one has to compute the full Hessian of the log-density $\log\pi(\vect{\theta}|\vect{y})$ to place the integration points around the mode. Finally, the mesh and the arguments $n_l,n_k$ of the inversion algorithm can be set dynamically every time when hyperparameter values change during optimization. However, this will most likely result in different workload for different values of $\vect{\theta}$ and requires one to re-generate and re-distribute the mesh at each iteration.

To sum up, the proposed method provides a scalable way to perform a Bayesian inference on large spatio-temporal GMRFs. Our approach is particularly well suited for daily weather applications, where thin 3D graphs produce compact separators in time. The source code with examples can be found on Github \url{https://github.com/abylayzhumekenov/parsinv}.

% references
\bibliographystyle{unsrtnat}
\bibliography{template}  %%% Uncomment this line and comment out the ``thebibliography'' section below to use the external .bib file (using bibtex) .

@article{bolin2020rational,
  title={The rational SPDE approach for Gaussian random fields with general smoothness},
  author={Bolin, David and Kirchner, Kristin},
  journal={Journal of Computational and Graphical Statistics},
  volume={29},
  number={2},
  pages={274--285},
  year={2020},
  publisher={Taylor \& Francis}
}

@article{golub1994matrices,
  title={Matrices, moments and quadrature},
  author={Golub, Gene H and Meurant, G{\'e}rard},
  journal={Pitman Research Notes in Mathematics Series},
  pages={105--105},
  year={1994},
  publisher={Longman Scientific \& Technical}
}

@inproceedings{petsc-efficient,
  author    = {Satish Balay and William~D. Gropp and Lois Curfman McInnes and Barry~F. Smith},
  title     = {Efficient Management of Parallelism in Object Oriented Numerical Software Libraries},
  booktitle = {Modern Software Tools in Scientific Computing},
  editor    = {E. Arge and A.~M. Bruaset and H.~P. Langtangen},
  publisher = {Birkh{\"{a}}user Press},
  pages     = {163--202},
  year      = {1997}
}

@article{chow2014preconditioned,
  title={Preconditioned Krylov subspace methods for sampling multivariate Gaussian distributions},
  author={Chow, Edmond and Saad, Yousef},
  journal={SIAM Journal on Scientific Computing},
  volume={36},
  number={2},
  pages={A588--A608},
  year={2014},
  publisher={SIAM}
}

@article{erisman1975computing,
  title={On computing certain elements of the inverse of a sparse matrix},
  author={Erisman, AM and Tinney, WF},
  journal={Communications of the ACM},
  volume={18},
  number={3},
  pages={177--179},
  year={1975},
  publisher={ACM New York, NY, USA}
}

@article{frommer2021analysis,
  title={Analysis of probing techniques for sparse approximation and trace estimation of decaying matrix functions},
  author={Frommer, Andreas and Schimmel, Claudia and Schweitzer, Marcel},
  journal={SIAM Journal on Matrix Analysis and Applications},
  volume={42},
  number={3},
  pages={1290--1318},
  year={2021},
  publisher={SIAM}
}

@article{hutchinson1989stochastic,
  title={A stochastic estimator of the trace of the influence matrix for Laplacian smoothing splines},
  author={Hutchinson, Michael F},
  journal={Communications in Statistics-Simulation and Computation},
  volume={18},
  number={3},
  pages={1059--1076},
  year={1989},
  publisher={Taylor \& Francis}
}

@article{parker2012sampling,
  title={Sampling Gaussian distributions in Krylov spaces with conjugate gradients},
  author={Parker, Albert and Fox, Colin},
  journal={SIAM Journal on Scientific Computing},
  volume={34},
  number={3},
  pages={B312--B334},
  year={2012},
  publisher={SIAM}
}

@article{schenk2001pardiso,
  title={PARDISO: a high-performance serial and parallel sparse linear solver in semiconductor device simulation},
  author={Schenk, Olaf and G{\"a}rtner, Klaus and Fichtner, Wolfgang and Stricker, Andreas},
  journal={Future Generation Computer Systems},
  volume={18},
  number={1},
  pages={69--78},
  year={2001},
  publisher={Elsevier}
}

@article{schneider2003krylov,
  title={A Krylov subspace method for covariance approximation and simulation of random processes and fields},
  author={Schneider, Michael K and Willsky, Alan S},
  journal={Multidimensional systems and signal processing},
  volume={14},
  pages={295--318},
  year={2003},
  publisher={Springer}
}

@article{siden2018efficient,
  title={Efficient covariance approximations for large sparse precision matrices},
  author={Sid{\'e}n, Per and Lindgren, Finn and Bolin, David and Villani, Mattias},
  journal={Journal of Computational and Graphical Statistics},
  volume={27},
  number={4},
  pages={898--909},
  year={2018},
  publisher={Taylor \& Francis}
}

@article{simpson2008fast,
  title={Fast sampling from a Gaussian Markov random field using Krylov subspace approaches},
  author={Simpson, Daniel P and Turner, Ian W and Pettitt, Anthony N},
  year={2008}
}

@article{simpson2013scalable,
  title={Scalable iterative methods for sampling from massive Gaussian random vectors},
  author={Simpson, Daniel P and Turner, Ian W and Strickland, Christopher M and Pettitt, Anthony N},
  journal={arXiv preprint arXiv:1312.1476},
  year={2013}
}

@article{tang2012probing,
  title={A probing method for computing the diagonal of a matrix inverse},
  author={Tang, Jok M and Saad, Yousef},
  journal={Numerical Linear Algebra with Applications},
  volume={19},
  number={3},
  pages={485--501},
  year={2012},
  publisher={Wiley Online Library}
}

@article{ubaru2017fast,
  title={Fast estimation of tr(f(A)) via stochastic Lanczos quadrature},
  author={Ubaru, Shashanka and Chen, Jie and Saad, Yousef},
  journal={SIAM Journal on Matrix Analysis and Applications},
  volume={38},
  number={4},
  pages={1075--1099},
  year={2017},
  publisher={SIAM}
}

@article{verbosio2017enhancing,
  title={Enhancing the scalability of selected inversion factorization algorithms in genomic prediction},
  author={Verbosio, Fabio and De Coninck, Arne and Kourounis, Drosos and Schenk, Olaf},
  journal={Journal of computational science},
  volume={22},
  pages={99--108},
  year={2017},
  publisher={Elsevier}
}

@article{rue2009approximate,
  title={Approximate Bayesian inference for latent Gaussian models by using integrated nested Laplace approximations},
  author={Rue, H{\aa}vard and Martino, Sara and Chopin, Nicolas},
  journal={Journal of the Royal Statistical Society Series B: Statistical Methodology},
  volume={71},
  number={2},
  pages={319--392},
  year={2009},
  publisher={Oxford University Press}
}

@article{lindgren2011explicit,
  title={An explicit link between Gaussian fields and Gaussian Markov random fields: the stochastic partial differential equation approach},
  author={Lindgren, Finn and Rue, H{\aa}vard and Lindstr{\"o}m, Johan},
  journal={Journal of the Royal Statistical Society Series B: Statistical Methodology},
  volume={73},
  number={4},
  pages={423--498},
  year={2011},
  publisher={Oxford University Press}
}

@article{lindgren2015bayesian,
  title={Bayesian spatial modelling with R-INLA},
  author={Lindgren, Finn and Rue, H{\aa}vard},
  journal={Journal of statistical software},
  volume={63},
  number={19},
  year={2015},
  publisher={University of California at Los Angeles}
}

@article{rue2017bayesian,
  title={Bayesian computing with INLA: a review},
  author={Rue, H{\aa}vard and Riebler, Andrea and S{\o}rbye, Sigrunn H and Illian, Janine B and Simpson, Daniel P and Lindgren, Finn K},
  journal={Annual Review of Statistics and Its Application},
  volume={4},
  pages={395--421},
  year={2017},
  publisher={Annual Reviews}
}

@article{simpson2017penalising,
  title={Penalising model component complexity: A principled, practical approach to constructing priors},
  author={Simpson, Daniel and Rue, H{\aa}vard and Riebler, Andrea and Martins, Thiago G and S{\o}rbye, Sigrunn H},
  year={2017}
}

@article{van2023new,
  title={A new avenue for Bayesian inference with INLA},
  author={Van Niekerk, Janet and Krainski, Elias and Rustand, Denis and Rue, H{\aa}vard},
  journal={Computational Statistics \& Data Analysis},
  volume={181},
  pages={107692},
  year={2023},
  publisher={Elsevier}
}

@article{lindgren2022spde,
  title={The SPDE approach for Gaussian and non-Gaussian fields: 10 years and still running},
  author={Lindgren, Finn and Bolin, David and Rue, H{\aa}vard},
  journal={Spatial Statistics},
  volume={50},
  pages={100599},
  year={2022},
  publisher={Elsevier}
}

@article{lindgren2020diffusion,
  title={A diffusion-based spatio-temporal extension of Gaussian Mat$\backslash$'ern fields},
  author={Lindgren, Finn and Bakka, Haakon and Bolin, David and Krainski, Elias and Rue, H{\aa}vard},
  journal={arXiv preprint arXiv:2006.04917},
  year={2020}
}

@article{fattah2022approximate,
  title={Approximate bayesian inference for the interaction types 1, 2, 3 and 4 with application in disease mapping},
  author={Fattah, Esmail Abdul and Rue, Haavard},
  journal={arXiv preprint arXiv:2206.09287},
  year={2022}
}

@article{gaedke2023parallelized,
  title={Parallelized integrated nested Laplace approximations for fast Bayesian inference},
  author={Gaedke-Merzh{\"a}user, Lisa and van Niekerk, Janet and Schenk, Olaf and Rue, H{\aa}vard},
  journal={Statistics and Computing},
  volume={33},
  number={1},
  pages={25},
  year={2023},
  publisher={Springer}
}

@article{gaedke2024integrated,
  title={Integrated Nested Laplace Approximations for Large-Scale Spatiotemporal Bayesian Modeling},
  author={Gaedke-Merzh{\"a}user, Lisa and Krainski, Elias and Janalik, Radim and Rue, H{\aa}vard and Schenk, Olaf},
  journal={SIAM Journal on Scientific Computing},
  volume={46},
  number={4},
  pages={B448--B473},
  year={2024},
  publisher={SIAM}
}

@article{amestoy2001fully,
  title={A fully asynchronous multifrontal solver using distributed dynamic scheduling},
  author={Amestoy, Patrick R and Duff, Iain S and L'Excellent, Jean-Yves and Koster, Jacko},
  journal={SIAM Journal on Matrix Analysis and Applications},
  volume={23},
  number={1},
  pages={15--41},
  year={2001},
  publisher={SIAM}
}

@article{li2003superlu_dist,
  title={SuperLU\_DIST: A scalable distributed-memory sparse direct solver for unsymmetric linear systems},
  author={Li, Xiaoye S and Demmel, James W},
  journal={ACM Transactions on Mathematical Software (TOMS)},
  volume={29},
  number={2},
  pages={110--140},
  year={2003},
  publisher={ACM New York, NY, USA}
}

@article{whittle1954stationary,
  title={On stationary processes in the plane},
  author={Whittle, Peter},
  journal={Biometrika},
  pages={434--449},
  year={1954},
  publisher={JSTOR}
}

@article{whittle1963stochastic,
  title={Stochastic-processes in several dimensions},
  author={Whittle, Peters},
  journal={Bulletin of the International Statistical Institute},
  volume={40},
  number={2},
  pages={974--994},
  year={1963},
  publisher={INT STATISTICAL INSTITUTE 428 PRINSES BEATRIXLAEN, VOORBURG, NETHERLANDS}
}

@article{liu1986computational,
  title={Computational models and task scheduling for parallel sparse Cholesky factorization},
  author={Liu, Joseph WH},
  journal={Parallel computing},
  volume={3},
  number={4},
  pages={327--342},
  year={1986},
  publisher={Elsevier}
}

@book{saad2003iterative,
  title={Iterative methods for sparse linear systems},
  author={Saad, Yousef},
  year={2003},
  publisher={SIAM}
}

@article{menne2012overview,
  title={An overview of the global historical climatology network-daily database},
  author={Menne, Matthew J and Durre, Imke and Vose, Russell S and Gleason, Byron E and Houston, Tamara G},
  journal={Journal of atmospheric and oceanic technology},
  volume={29},
  number={7},
  pages={897--910},
  year={2012},
  publisher={American Meteorological Society}
}

@book{rue2005gaussian,
  title={Gaussian Markov random fields: theory and applications},
  author={Rue, Havard and Held, Leonhard},
  year={2005},
  publisher={Chapman and Hall/CRC}
}

@article{papandreou2010gaussian,
  title={Gaussian sampling by local perturbations},
  author={Papandreou, George and Yuille, Alan L},
  journal={Advances in Neural Information Processing Systems},
  volume={23},
  year={2010}
}

@article{chen1985lumped,
  title={The lumped mass finite element method for a parabolic problem},
  author={Chen, Chuan Miao and Thom{\'e}e, Vidar},
  journal={The ANZIAM Journal},
  volume={26},
  number={3},
  pages={329--354},
  year={1985},
  publisher={Cambridge University Press}
}

@article{amari1993backpropagation,
  title={Backpropagation and stochastic gradient descent method},
  author={Amari, Shun-ichi},
  journal={Neurocomputing},
  volume={5},
  number={4-5},
  pages={185--196},
  year={1993},
  publisher={Elsevier}
}

@article{kiwiel2001convergence,
  title={Convergence and efficiency of subgradient methods for quasiconvex minimization},
  author={Kiwiel, Krzysztof C},
  journal={Mathematical programming},
  volume={90},
  pages={1--25},
  year={2001},
  publisher={Springer}
}

@article{bottou1998online,
  title={Online algorithms and stochastic approximations},
  author={Bottou, L{\'e}on},
  journal={Online learning in neural networks},
  year={1998},
  publisher={Cambridge University Press}
}





% \end{thebibliography}

\end{document}